\documentclass[12pt,a4paper,onecolumn,draftcls]{IEEEtran}
\usepackage{epsfig,graphicx,subfigure,psfrag,amsmath,cases}
\usepackage{latexsym,amssymb,amsmath,epsfig,subfigure,algorithm,mathtools,bm,relsize}
\usepackage{algorithmic}
\usepackage{color}
\usepackage{url}
\usepackage{scrtime}
\usepackage{array}

\usepackage{epstopdf}

 \newcommand{\qed}{\hfill \ensuremath{\blacksquare}}

\author{\normalsize \IEEEauthorblockN{Elena Boshkovska\IEEEauthorrefmark{1}, Derrick Wing Kwan Ng\IEEEauthorrefmark{2}, Linglong Dai\IEEEauthorrefmark{3},  and Robert Schober\IEEEauthorrefmark{1}\\ \thanks{ This paper has been  accepted in part for presentation at  IEEE Globecom 2017 \cite{CN:Elena_WPC_HWI}.}
}
\IEEEauthorblockA{\IEEEauthorrefmark{1}Friedrich-Alexander-University Erlangen-N\"urnberg (FAU), Germany\\}

 \IEEEauthorblockA{\IEEEauthorrefmark{2}The University of New South Wales, Australia
    \\}
 \IEEEauthorblockA{\IEEEauthorrefmark{3}Tsinghua University, Beijing, China
    \\}

}

\title{Power-Efficient and Secure WPCNs with Hardware Impairments and  Non-Linear EH Circuit}

\newtheorem{Thm}{Theorem}
\newtheorem{Lem}{Lemma}

\newtheorem{Remark}{Remark}
\newcommand{\abs}[1]{\lvert#1\rvert}

\newcommand{\norm}[1]{\lVert#1\rVert}
\DeclareMathOperator{\Tr}{\mathrm{Tr}}
\DeclareMathOperator{\zero}{\mathbf{0}}
\DeclareMathOperator{\Rank}{\mathrm{Rank}}

\DeclareMathOperator{\diag}{\mathrm{diag}}

\DeclareMathOperator{\mino}{\mathrm{minimize}}

\DeclareMathAlphabet\mathbfcal{OMS}{cmsy}{b}{n}

\newcolumntype{L}{>{\arraybackslash\raggedright}m{7cm}}
\makeatletter

\newcommand{\Rmnum}[1]{\expandafter\@slowromancap\romannumeral #1@}
\makeatother

\begin{document}

\maketitle
\vspace*{-19mm}
\begin{abstract}
In this paper, we design a robust resource allocation algorithm for a wireless-powered communication network (WPCN) taking into account residual hardware impairments (HWIs) at the transceivers, the imperfectness of the channel state information,  and the non-linearity of practical radio frequency energy harvesting circuits. In order to ensure power-efficient secure communication, physical layer  security techniques are exploited to deliberately degrade the channel quality of a multiple-antenna eavesdropper. The resource allocation algorithm design is formulated as a non-convex optimization problem for minimization of the total consumed power in the network, while guaranteeing the quality of service of the information receivers in terms of secrecy rate. The globally optimal solution of the optimization problem is obtained via a two-dimensional search and semidefinite programming relaxation. To strike a balance between computational complexity and system performance, a low-complexity iterative  suboptimal resource allocation algorithm is then proposed.
 Numerical results demonstrate that both the proposed optimal and suboptimal schemes can significantly reduce the total system power consumption required for guaranteeing secure communication, and   unveil
the impact of HWIs on the system performance: (1) residual HWIs create a system performance bottleneck in the high transmit/receive power regimes; (2) increasing the number of transmit antennas  can effectively reduce the system power consumption and alleviate the performance degradation due to residual  HWIs; (3) imperfect CSI increases the system power consumption and exacerbates the impact of residual HWIs.
\end{abstract}%\vspace*{-6mm}
%\begin{keywords} Hardware impairments, wireless-powered communication, non-linear energy harvesting model, energy beamforming, physical layer security.
%\end{keywords}
\newpage
%
%\large\normalsize
\section{Introduction}

\label{sect1}

Wireless charging of battery-powered devices in wireless communication networks via wireless power transfer (WPT) technology could prolong the lifetime of the networks. In fact, the concept of wireless-powered communication networks (WPCNs), where wireless devices are powered via radio frequency (RF) electromagnetic waves, has gained considerable attention recently in the context of enabling sustainability via WPT \cite{JR:WPCN_Overview_new}. In particular, it is expected that the number of interconnected wireless devices will increase to up to $50$ billion by 2020 \cite{MG:IoT}, due to the roll-out of  the Internet-of-Things (IoT).  A large portion of these wireless devices, some of which may be inaccessible for frequent battery replacement, could be powered wirelessly by dedicated power stations via RF-based WPT technology to facilitate their information transmissions \cite{JR:QQ_WPCN}\nocite{JR:Rui_WPCN}--\cite{JR:He_WPCN}. Specifically, RF-based WPT offers a more stable and controllable source of energy compared to natural energy sources, such as solar, wind, and tidal, etc., which are usually climate and location dependent \cite{Krikidis2014}--\nocite{JR:Xiaoming_magazine_SWIPT}\cite{Ding2014}. More importantly, RF-based WPT exploits the broadcast nature of the wireless medium which enables one-to-many simultaneous long-range wireless charging. On the other hand, the large number of wireless devices in future networks encourages the use of low-quality and low-cost hardware components in order to reduce deployment costs. However, RF transceivers equipped with cheap hardware components suffer from various kinds of hardware impairments (HWIs) resulting potentially in a performance degradation for communications. These HWIs are caused by non-linear power amplifiers, frequency and phase offsets, in-phase and quadrature (I/Q) imbalance, and quantization noise. Although the negative impact of HWIs on the system performance can be reduced by calibration and compensation algorithms, residual distortions at the transceivers that depend on the power of the transmitted/received signal are inevitable \cite{coord_beamforming_impair}\nocite{JR:Jiayi_HWI,MIMO_res_impair}--\cite{book:Emil_HWI}. Hence, existing resource allocation algorithms for multiuser WPCNs, e.g. \cite{JR:QQ_WPCN}--\cite{JR:He_WPCN}, designed based on the assumption of ideal hardware, may lead to substantial performance losses in practical systems.

The increasing number of wireless devices also poses a threat to communication security in future wireless networks due to the enormous amount of data transmitted over wireless channels  \cite{JR:Secrecy_WIPT_Magazine}\nocite{JR:Yuan_Secure_SWIPT}--\cite{L:Secure_WPCN}. Nowadays, wireless communication security is ensured by cryptographic encryption algorithms operating in the application layer. Unfortunately, these traditional security methods may not be applicable in future wireless networks with large numbers of transceivers, since encryption algorithms usually require secure secret key distribution and management via an authenticated third party. Recently, physical layer (PHY) security has been proposed as an effective complementary technology to the existing encryption algorithms for providing secure communication \cite{JR:Secrecy_WIPT_Magazine}--\nocite{JR:Yuan_Secure_SWIPT,L:Secure_WPCN,JR:SWIPT_DAS,JR:WP_massive_MIMO}\cite{JR:WP_friendly_jammer}. Specifically, PHY security exploits the unique characteristics of wireless channels, such as fading, noise, and interference, to protect the communication between legitimate devices from eavesdropping.
In this context, the authors of \cite{L:Secure_WPCN} designed a resource allocation algorithm that jointly optimizes the transmit power, the duration of WPT, and the direction of spatial beams to facilitate security in WPCNs. In \cite{JR:SWIPT_DAS}, beamforming design was studied for secrecy provisioning in distributed antenna systems with WPT.  The authors of \cite{JR:WP_massive_MIMO} investigated the design of secure transmission in wireless-powered relaying systems.  In \cite{JR:WP_friendly_jammer}, the use of a wireless-powered
friendly jammer was proposed to enable secure communication in a point-to-point communication system. However, most of the existing works on secure WPT systems were based on the assumption of ideal hardware \cite{L:Secure_WPCN}--\cite{JR:WP_friendly_jammer} and are not applicable to practical systems with HWIs. Recently, the notion of secure communication under the consideration of HWIs has been pursued. For instance, the work in \cite{JR:HW_massive_MIMO_secure} considered the analysis and design of secure massive multiple-input multiple-output (MIMO) systems in the presence of a passive eavesdropper and HWIs at the transceivers. Besides, the authors of \cite{RF_impair} studied the impact of residual HWIs on the performance of  a two-way WPT-based cognitive relay network, where the relay is powered by harvesting energy from the signals transmitted by the source in the RF.  In \cite{JR:Vincent_HWI_SWIPT}, the authors analyzed the impact  of  phase  noise  on  downlink  WPT  in  secure  multiple antennas systems. However, the authors  of \cite{RF_impair,JR:Vincent_HWI_SWIPT} assumed an overly simplified linear  energy harvesting (EH) model for the end-to-end WPT characteristic. Yet, measurements of practical RF-based EH circuits demonstrate a highly non-linear end-to-end WPT characteristic \cite{CN:EH_measurement_2}, which implies that transmission schemes and algorithms designed based on the conventional linear EH model may cause performance degradation in practical implementations. Moreover, the transmission strategies in \cite{JR:HW_massive_MIMO_secure}--\cite{JR:Vincent_HWI_SWIPT}  were not optimized. Hence, the design of resource allocation for secure communication in WPCNs with the non-linear EH circuits suffering from HWIs is an important open problem.

To address the above issues, we propose a resource allocation algorithm design, which aims at  providing power-efficient and secure communication in WPCNs in the presence of a multiple-antenna eavesdropper.  The resource allocation algorithm design is formulated as a non-convex optimization problem taking into account the non-linearity of the EH circuits, the existence of residual HWIs at the transceivers, and the imperfectness of the  channel state information  (CSI) of the eavesdropper. We  minimize the total consumed power while guaranteeing the quality of service (QoS) at the information receivers (IRs) in the WPCN. The optimal solution of the proposed problem is obtained via a two-dimensional search and semidefinite programming (SDP) relaxation. The proposed solution unveils that information beamforming from the access point in the direction of the IRs is optimal and that the SDP relaxation is tight. Besides, a low-computational complexity iterative suboptimal scheme is proposed to obtain a suboptimal solution.  Numerical results demonstrate that the proposed schemes can significantly reduce the power consumption in the considered WPCN compared to two baseline schemes.

\section{System Model}
\label{sec:system_model}
In this section, we first present some notations and the considered system model. Then, we discuss the energy harvesting and hardware impairment models adopted for power-efficient resource allocation algorithm design.
\subsection{Notation}
We use boldface capital and lower case letters to denote matrices and vectors, respectively. $\mathbf{A}^H$, $\Tr(\mathbf{A})$, $\det(\mathbf{A})$, $\mathbf{A}^{-1}$, $\Rank(\mathbf{A})$, and  $\lambda_{\max}(\mathbf{A})$ represent the  Hermitian transpose, trace, determinant, inverse, rank, and  maximum eigenvalue of  matrix $\mathbf{A}$, respectively; $\mathbf{A}\succeq \mathbf{0}$ indicates that $\mathbf{A}$ is a  positive semidefinite matrix;   $\mathbf{I}_{N}$
denotes the $N\times N$ identity matrix.  $\mathbb{C}^{N\times M}$ denotes the space of all $N\times M$ matrices with complex entries.
$\mathbb{H}^N$ represents the set of all $N$-by-$N$ complex Hermitian matrices.
 $|{\cdot}|$ and $\norm{\cdot}_{\mathrm{F}}$ represent the absolute value of a complex scalar and the Frobenius norm, respectively. The distribution of a circularly symmetric complex Gaussian (CSCG) vector with mean vector $\mathbf{x}$ and covariance matrix $\mathbf{\Sigma}$  is denoted by ${\cal CN}(\mathbf{x},\mathbf{\Sigma})$, and $\sim$ means ``distributed as".  $\cal E\{\cdot\}$ denotes statistical expectation. $[x]^{+}$ stands for $\max \{0, x\}$. $\nabla_{\mathbf{x}}f(\mathbf{x})$ represents the partial derivative of function $f(\mathbf{x})$ with respect to the elements of vector $\mathbf{x}$. Furthermore, $\diag[\mathbf{x}]$ is a diagonal matrix with the elements of $\mathbf{x}$ on the main diagonal. $[\mathbf{A}]_{n,n}$ returns the element in the $n$-th row and $n$-th column of square matrix $\mathbf{A}$.  $\mathbf{S}_m$ is a square matrix with all entries equal to $0$ except for the $m$-th diagonal element which is equal to $1$.
\subsection{System Model}
\begin{figure}[t]
\centering
\includegraphics[width=4.5 in]{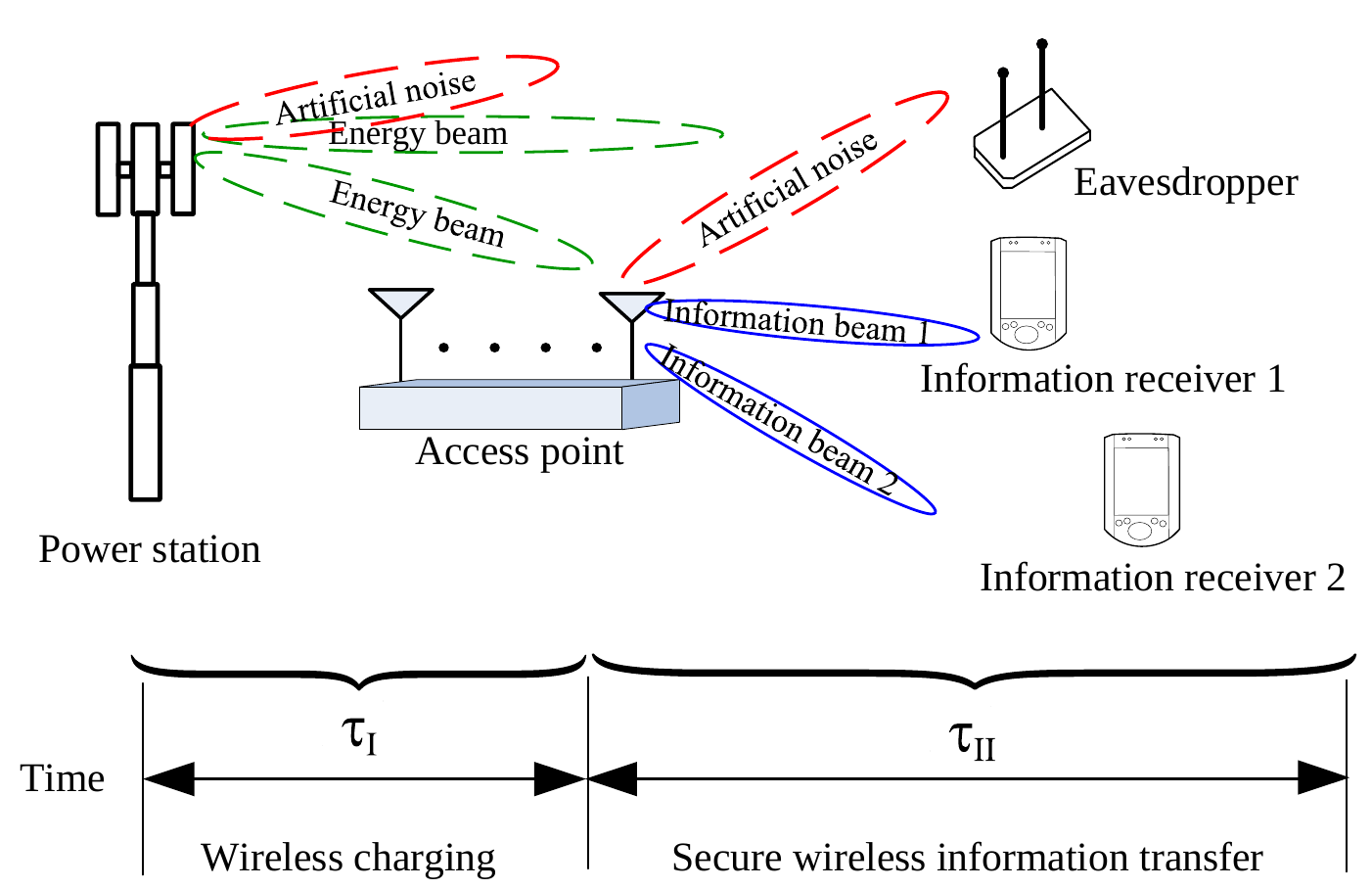}
\caption{A WPCN employing two transmission phases with $K=2$ information receivers (IRs) and one multiple-antenna eavesdropper.}
\label{fig:system_model}\vspace*{-6mm}
\end{figure}

We focus on a WPCN which consists of a power station\footnote{In this work, we assume that the PS is connected to the main grid with a continuous and stable energy supply.} (PS), an access point (AP), $K$ IRs, and one eavesdropper (Eve), cf. Figure \ref{fig:system_model}. We assume that the PS, the AP, and Eve are equipped with $N_{\mathrm{PS}}\geq1$, $N_{\mathrm{AP}}\geq1$, and $N_{\mathrm{EV}}\geq1$ antennas\footnote{We note that an eavesdropper equipped with $N_{\mathrm{EV}}$ antennas is equivalent to multiple eavesdroppers with a total of $N_{\mathrm{EV}}$ antennas which are connected to a joint processing unit performing cooperative eavesdropping. Besides, we assume $N_{\mathrm{PS}}+N_{\mathrm{AP}}\geq N_{\mathrm{EV}}$ to enable secure communication.}, respectively. The IRs are single-antenna devices for hardware simplicity. The communication in the WPCN comprises two transmission phases as shown in Figure \ref{fig:system_model}. We assume that the fading channels in both phases are frequency flat and slowly time-varying. In particular, Phase I, with a time duration of $\tau_{\mathrm{I}}$, is reserved for wireless charging, where the PS transmits a dedicated energy beam to the energy-constrained AP.
The instantaneous received signal at the AP during Phase I is given by
\begin{eqnarray}\label{eqn:signal_at_AP}
\mathbf{y}_{\mathrm{AP}}&=&\mathbf{L}^H\Big(\mathbf{v}+\bm{\xi}^{(t)}\Big)+\bm{\xi}^{(r)} + \mathbf{n}_{\mathrm{AP}},
\end{eqnarray}
where  $\mathbf{v}\in\mathbb{C}^{N_{\mathrm{PS}}\times1}$ is the energy signal vector adopted in Phase I for wireless charging with covariance matrix $\mathbf{V} = \mathcal{E} \{ \mathbf{v}\mathbf{v}^H \}$. The channel matrix between the PS and the AP is denoted by  $\mathbf{L}\in\mathbb{C}^{N_{\mathrm{PS}}\times N_{\mathrm{AP}}}$ and captures the joint effect of path loss and multipath fading. Vector $\mathbf{n}_{\mathrm{AP}}\sim{\cal CN}(\zero,\sigma_{n}^2\mathbf{I}_{N_{\mathrm{AP}}})$ represents the additive white Gaussian noise (AWGN) at the AP where $\sigma_{n}^2$ denotes the noise variance at each antenna of the AP. In \eqref{eqn:signal_at_AP}, $\bm{\xi}^{(t)}$ $\in \mathbb{C}^{N_{\mathrm{PS}} \times 1}$ and $\bm{\xi}^{(r)}$ $\in \mathbb{C}^{N_{\mathrm{AP}} \times 1}$ represent the random residual HWIs after compensation introduced at the transmitter and receiver during Phase I, respectively. The model adopted for the residual HWIs will be presented later in the next section.

In Phase II, for a time duration of $\tau_{\mathrm{II}}$, the AP transmits $K$ independent signals to the $K$ IRs simultaneously. Because of the broadcast nature of wireless channels, there is a security threat due to potential eavesdropping. To circumvent this threat, both the AP and the PS deliberately emit artificial noise to degrade the channel quality of the eavesdropper \cite{JR:Secrecy_WIPT_Magazine}. Therefore, the instantaneous received signal at IR $k$ in Phase II is given by
\begin{equation}\label{eqn:signal_at_IR}
y_{\mathrm{IR}_k} = \mathbf{h}_k^H\bigg(\sum_{i=1}^{K}\mathbf{w}_i s_i+\underbrace{\mathbf{u} + \bm{\varsigma}^{(t)}}_{\text{Jamming from AP}}\bigg)+\mathbf{f}_k^H \underbrace{\Big( \mathbf{z}+\bm{\kappa}^{(t)}\Big)}_{\text{Jamming from PS}}+\varsigma_k^{(r)} + n_{\mathrm{IR}_k},
\end{equation}
 where $s_k \in \mathbb{C}$ and $\mathbf{w}_k \in \mathbb{C}^{N_{\mathrm{AP}} \times 1}$ are the information symbol for IR $k$ and the corresponding beamforming vector, respectively.  Without loss of generality, we assume that $\mathcal{E}\{|s_k|^2\}=1, \forall k$. $\mathbf{h}_k \in\mathbb{C}^{N_{\mathrm{AP}}\times 1}$ is the channel vector between the AP and IR $k$, while $\mathbf{f}_k \in\mathbb{C}^{N_{\mathrm{PS}}\times 1}$ denotes the channel vector between the PS and IR $k$. Furthermore, $\mathbf{u} \sim {\cal CN}(\mathbf{0}, \mathbf{U})$ and $\mathbf{z} \sim {\cal CN}(\mathbf{0}, \mathbf{Z})$ are the  Gaussian pseudo-random energy signal sequences   broadcasted, i.e., the artificial noise,  by the AP and the PS, respectively, where $\mathbf{U}\in \mathbb{H}^{N_{\mathrm{AP}}},\mathbf{U}\succeq \mathbf{0},$ and  $\mathbf{Z}\in \mathbb{H}^{N_{\mathrm{PS}}},\mathbf{Z}\succeq \mathbf{0}$, denote the corresponding covariance matrices, respectively. These two noise processes are exploited by the AP and the PS to degrade the channel quality of the eavesdropper via jamming.  $\bm{\varsigma}^{(t)} \in \mathbb{C}^{N_{\mathrm{AP}}\times 1}$ and    $\bm{\kappa}^{(t)}$ $\in \mathbb{C}^{N_{\mathrm{PS}} \times 1}$ represent the random residual transmitter HWIs after compensation introduced by the AP and the PS in Phase II, respectively, while $\varsigma_k^{(r)} \in \mathbb{C}$ represents the residual receiver HWIs introduced by IR $k$. $n_{\mathrm{IR}_k} \sim {\cal CN} (0, \sigma_{\mathrm{IR}_k}^2)$ is the AWGN at IR $k$ with noise power $\sigma_{\mathrm{IR}_k}^2.$

 The instantaneous received signal at Eve in Phase II is given by
\begin{equation}\label{eqn:signal_at_EVE}
\mathbf{y}_{\mathrm{E}}=\mathbf{G}^H\bigg(\sum_{i=1}^{K}\mathbf{w}_i s_i + \underbrace{\mathbf{u}+\bm{\varsigma}^{(t)}}_{\text{Jamming from AP}}\bigg) + \mathbf{E}^H\bigg(\underbrace{\mathbf{z}+\bm{\kappa}^{(t)}}_{\text{Jamming from PS}}\bigg)+ \mathbf{n}_{\mathrm{E}},
\end{equation}
where $\mathbf{G} \in \mathbb{C}^{N_{\mathrm{AP}}\times N_{\mathrm{E}}}$ and $\mathbf{E} \in \mathbb{C}^{N_{\mathrm{PS}}\times N_{\mathrm{E}}}$  denote the channel matrices of the AP-to-Eve links and the PS-to-Eve links, respectively.   $\mathbf{n}_{\mathrm{E}}\sim{\cal CN}(\zero,\sigma_{\mathrm{E}}^2\mathbf{I}_{N_{\mathrm{EV}}})$ is the AWGN vector at the eavesdropper with noise power $\sigma_{\mathrm{E}}^2$.
 In this work, we assume that ideal hardware is available at the eavesdropper, i.e., there are no HWIs at the eavesdropper,  which constitutes the worst case for communication security.

\subsection{Hardware Impairment Model}
In this paper, we adopt the general HWI model proposed in \cite[Chapter 4]{book:Emil_HWI}, \cite[Chapter 7]{book:impair}. In particular, the residual distortion caused by the aggregate effect of different HWIs, such as I/Q imbalance, phase noise, and power amplifier non-linearities is modeled as a Gaussian random variable whose variance scales with the power of the signals transmitted and received  at the transmitter and the receiver, respectively. This model has been widely used in the literature to study the impact of transceiver HWIs on the performance of communication systems \cite{coord_beamforming_impair}, \cite{MIMO_res_impair}, \cite{RF_impair}. Besides, the authors of \cite{MIMO_res_impair} showed that this model accurately captures the residual distortions caused by the joint effect of various HWIs in practical multiple-antenna systems.

 \begin{figure}[t]
\centering
\includegraphics[width=4.5 in]{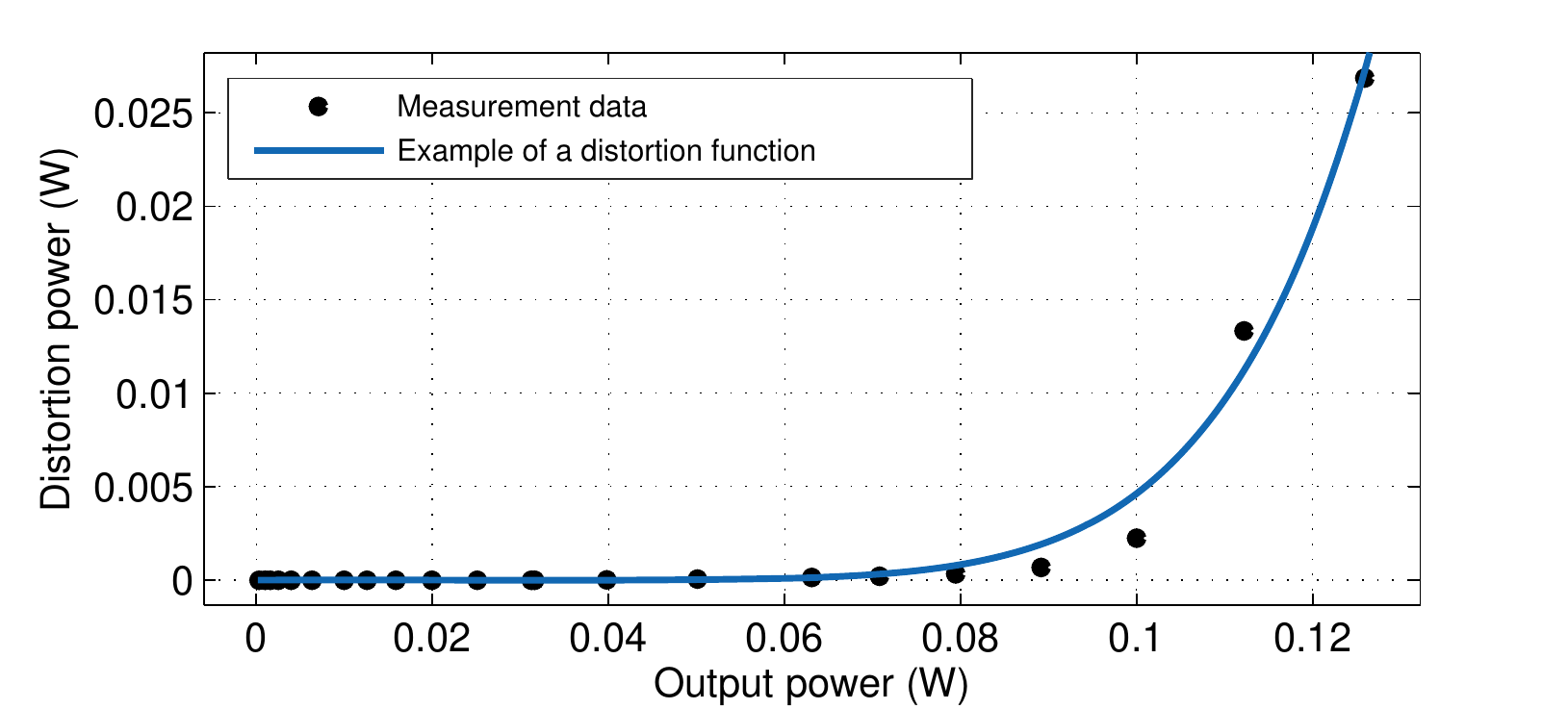}
\caption{ A comparison between the convex distortion model  in \eqref{eta_func} and measurement data
for a long term evolution (LTE) transmitter power amplifier \cite{tech_power_amplifier}. In this example, parameters $k_{\mathrm{1}}=2.258\times 10^{5}$ and $k_{\mathrm{2}}=7.687$ in \eqref{eta_func} are calculated by a standard curve fitting tool.}.\vspace*{-6mm}
\label{fig:convex-increasing-distortion}
\end{figure}
Hence, the distortion noises caused by the transmitter HWIs  at the PS in Phase I and Phase II are modeled as $\bm{\xi}^{(t)} \sim {\cal CN}(\mathbf{0}, \mathbf{\Phi})$ and $\bm{\kappa}^{(t)} \sim {\cal CN}(\mathbf{0}, \mathbf{\Theta})$, respectively. $\mathbf{\Phi} \in \mathbb{C}^{N_{\mathrm{PS}}\times N_{\mathrm{PS}}}$ and $\mathbf{\Theta} \in \mathbb{C}^{N_{\mathrm{PS}}\times N_{\mathrm{PS}}}$ are diagonal covariance matrices which contain on their main diagonal the distortion noise variances at each antenna of the PS  in Phase I and Phase II, respectively, and  are given by
\begin{eqnarray}\label{eqn:PS_transmit_cov}
\mathbf{\Phi} &=& \diag{\bigg[\eta_1\bigg({P_{\mathrm{av},1}^{\mathrm{PS-I}}}\bigg), \ldots, \eta_{N_{\mathrm{PS}}}\bigg({P_{\mathrm{av},N_{\mathrm{PS}}}^{\mathrm{PS-I}}}\bigg)\bigg]} \,\,\mbox{and}\\
\mathbf{\Theta} &=& \diag{\bigg[\eta_1\bigg({P_{\mathrm{av},1}^{\mathrm{PS-II}}}\bigg), \ldots, \eta_{N_{\mathrm{PS}}}\bigg({P_{\mathrm{av},N_{\mathrm{PS}}}^{\mathrm{PS-II}}}\bigg)\bigg]}. \label{eqn:PS_transmit_cov2}
\end{eqnarray}
In \eqref{eqn:PS_transmit_cov} and \eqref{eqn:PS_transmit_cov2}, $P_{\mathrm{av},m}^{\mathrm{PS-I}} = \mathcal{E}\{\norm{\mathbf{S}_m\mathbf{v}}_{\mathrm{F}}^2\} = {[\mathbf{V}]_{m,m}}$ and $P_{\mathrm{av},m}^{\mathrm{PS-II}} = \mathcal{E}\{\norm{\mathbf{S}_m\mathbf{z}}_{\mathrm{F}}^2\} = {[\mathbf{Z}]_{m,m}}$, $m \in \{1,\ldots,N_{\mathrm{PS}}\}$, respectively,  are the average powers of the transmit signal at the $m$-th antenna of the PS in Phase I and Phase II, respectively. Also,
$\eta_{m}(\cdot), \forall m$, is a convex, continuous, and monotonically increasing distortion function which quantifies the impact of the HWIs for a given average power of the transmit signal at the $m$-th antenna\footnote{Note that  the model considered here directly maps the signal power to the distortion power while the one proposed in \cite[Chapter 4]{book:Emil_HWI} maps the signal magnitude to the distortion magnitude. }, i.e., the function maps the average power of the signal to a specific distortion value. For example, the transmitter distortion function can be modeled by the following convex increasing function:
\begin{equation}\label{eta_func}
\eta_m(x_m) =k_{\mathrm{1}} x_m^{k_{\mathrm{2}}}  \quad [{\text{Watt}}],
\end{equation}
where $x_m$ [Watt] is  the average transmit power at antenna $m$. Constants $k_{\mathrm{1}}\geq 0$ and $k_{\mathrm{2}}\geq 1$ are model parameters which are chosen such that they fit the measurements of the associated practical systems\footnote{In practice, the value of the distortion function of the transmitter HWIs usually grows at least linearly with respect to the transmit power which leads to $k_{\mathrm{2}}\geq 1$.}.
In Figure \ref{fig:convex-increasing-distortion}, we illustrate that the proposed model for the transmitter distortion function  in \eqref{eta_func} closely matches the experimental results in \cite{tech_power_amplifier}.

Similarly, the transmitter HWIs at the AP in Phase II are modeled as $\bm{\varsigma}^{(t)}\sim {\cal CN}(\mathbf{0}, \mathbf{\Psi})$ with covariance matrix
%\mathcal{E}\{\norm{\mathbf{S}_{[n]}(\sum_{k=1}^{K}\mathbf{w}_k s_k+\mathbf{v})}_{\mathrm{F}}\}
\begin{equation}\label{eqn:AP_transmit_cov}
\mathbf{\Psi} = \diag{ \bigg[\eta_1\bigg({P_{\mathrm{av},1}^{\mathrm{AP}}}\bigg), \ldots, \eta_{N_{\mathrm{AP}}}\bigg({P_{\mathrm{av},{N_{\mathrm{AP}}}}^{\mathrm{AP}}}\bigg) \bigg]}, \,
\end{equation}
where $P_{\mathrm{av},n}^{\mathrm{AP}} = \mathcal{E}\Big\{\norm{\mathbf{S}_n(\sum_{k=1}^{K}\mathbf{w}_k s_k+\mathbf{u})}_{\mathrm{F}}^2\Big\} = {\Big[ \sum_{k=1}^{K} \mathbf{w}_k\mathbf{w}_k^H+ \mathbf{U} \Big]}_{n,n}$, $\forall n \in \{1,\ldots, $ $N_{\mathrm{AP}}\}$. On the other hand, the received signal is affected mostly by phase noises and I/Q imbalances \cite[Chapter 7]{book:impair}. In the considered WPCN, these residual HWIs are modeled by the receiver distortion noise at the AP, $\bm{\xi}^{(r)} \in \mathbb{C}^{N_{\mathrm{AP}}\times 1}$, cf. \eqref{eqn:signal_at_AP}, where $\bm{\xi}^{(r)} \sim {\cal CN} (\mathbf{0}, \sigma^2_{d_{\mathrm{AP}}} \mathbf{I}_{N_{\mathrm{AP}}})$. Moreover, $\sigma_{d_{\mathrm{AP}}}^2 = \nu\Big({\mathcal{E}\{\norm{\mathbf{L}^H\mathbf{v}}_{\mathrm{F}}^2}\}\Big)= \nu\Big(\Tr(\mathbf{V}\mathbf{L}\mathbf{L}^H)\Big)$, where $\nu(\cdot)$ is a convex, continuous, and monotonically increasing function that models the receiver impairment characteristic. Additionally, the receiver HWIs at each IR during Phase II are given by $\varsigma_k^{(r)} \sim {\cal CN}(0, \sigma_{d_{\mathrm{IR}_k}}^2)$, with $\sigma_{d_{\mathrm{IR}_k}}^2 = \nu\Big({\mathcal{E}\{\norm{\mathbf{h}_k^H(\sum_{i=1}^{K}\mathbf{w}_i s_i +\mathbf{u})+\mathbf{f}_k^H\mathbf{z}}_{\mathrm{F}}^2}\}\Big)=\nu
\Big(\Tr(\mathbf{H}_k(\mathbf{U}+\sum_{i=1}^{K}\mathbf{w}_i\mathbf{w}_i ^H )+\mathbf{F}_k\mathbf{Z})\Big)$, where $\mathbf{F}_k=\mathbf{f}_k\mathbf{f}_k^H$ and $\mathbf{H}_k=\mathbf{h}_k\mathbf{h}_k^H$  are introduced for notational simplicity.
According to \cite{coord_beamforming_impair},  a suitable choice of the receiver distortion function is  $\nu(x) = (\frac{k_\mathrm{3}}{100})^2x$ [Watt], where $x$ is the average power of the received signal and $k_{\mathrm{3}}$ is a constant model parameter with a typical range of $k_{\mathrm{3}}\in[0,\,15]$.

\subsection{Energy Harvesting Model}
In the considered WPCN, we exploit the energy and  artificial noise signals transmitted by the PS in Phase I and Phase II to charge the AP and to facilitate secure information transfer, respectively. In this paper, we adopt the practical non-linear RF-based EH model proposed in  \cite{Letter_non_linear} to characterize the end-to-end WPT at the AP. The total energy harvested by the AP in Phase I is given by
 \begin{eqnarray}\label{eqn:EH_non_linear}
 \Xi_{\mathrm{tot}} (\omega) &=& \frac{\frac{M}{1+\exp(-a(\omega-b))}  -M\Omega}{1 - \Omega},\quad \Omega = \frac{1}{1+\exp(ab)},\\
 \omega &=& \Tr\Big(\mathbf{L}^H(\mathbf{V}+\mathbf{\Phi})\mathbf{L}\Big),\notag
 \end{eqnarray}
where $\omega$ represents the received RF power at the AP. The parameters $M$, $a$, and $b$ in \eqref{eqn:EH_non_linear} capture the joint effects of various non-linear phenomena caused by hardware limitations in practical EH circuits. More specifically, $M$ represents the maximum power that can be harvested by the EH circuit, as the circuit becomes saturated for exceedingly large received RF powers. Moreover, $a$ and $b$ depend on several physical hardware phenomena, such as circuit sensitivity  and potential current leakage. In fact, the adopted non-linear EH model was shown to accurately characterize the behavior of various practical EH circuits \cite{Letter_non_linear,Journal_non_linear}. In contrast, the conventional linear EH model, which is widely used in the literature \cite{L:Secure_WPCN}--\cite{JR:WP_friendly_jammer,RF_impair}, may lead to  performance degradation due to a severe model mismatch for resource allocation algorithm design.

 \begin{Remark}In practice,
the EH hardware circuit of the AP is fixed and  parameters $a,b,$ and $M$ of the non-linear model in \eqref{eqn:EH_non_linear} can be determined by  a  standard  curve  fitting  tool. \end{Remark}

\subsection{Channel State Information}
%%%%%%%%%%%%%%%%%%%%%%%%%%%%%%%%%%%%%%%%%%%%%%%%%%%%%
In practice, handshaking is performed between the legitimate PS, the AP, and the $K$ IRs. As a result, accurate CSI can be obtained by exploiting the pilot sequences embedded in the handshaking signals.  In this paper, we assume that the CSI of  $\mathbf{L}$, $\mathbf{h}_k$, and $\mathbf{f}_k$ is available for resource allocation algorithm design.
    In contrast, the potential eavesdropper may not directly interact with the transmitters and it is difficult to obtain  perfect CSI for the corresponding links.   To capture the impact of imperfect CSI knowledge of the eavesdropper's channels on the system performance, we adopt the deterministic model from \cite{Journal_non_linear}--\nocite{JR:MOOP_robust_secure_FD,JR:Robust_error_models1}\cite{JR:secrecy_capacity_prob_det_models}. To this end, the  CSI of the relevant communication links is modeled as
\begin{eqnarray}\label{eqn:outdated_CSI-set1}
\hspace*{-6mm}\mathbf{G}\hspace*{-2mm}&=&\hspace*{-2mm}\mathbf{\widehat G} + \Delta\mathbf{G},\quad   {\bm\Xi }^{\mathrm{AP}}_\mathrm{E}\triangleq \Big\{\Delta\mathbf{G}\in \mathbb{C}^{N_{\mathrm{AP}}\times {N_{\mathrm{E}}}}  :\norm{\Delta\mathbf{G}}_\mathrm{F}  \le \upsilon_\mathrm{AP\rightarrow E}\Big\},\, \quad \\
\hspace*{-6mm}\mathbf{E}&=&\hspace*{-2mm}\mathbf{\widehat E} + \Delta\mathbf{E},\,\,\,\quad {\bm\Xi }^{\mathrm{PS}}_\mathrm{E}\triangleq \Big\{\Delta\mathbf{E}\in \mathbb{C}^{N_{\mathrm{PS}}\times {N_{\mathrm{E}}}}  :\norm{\Delta\mathbf{E}}_\mathrm{F}  \le \upsilon_\mathrm{PS\rightarrow E}\Big\}, \label{eqn:outdated_CSI-set2}
% \hspace*{-6mm} \mathbf{L}\hspace*{-2mm}&=&\hspace*{-2mm}\mathbf{\widehat L} + \Delta\mathbf{L},\,\,\,\quad {\bm\Xi }^{\mathrm{PS}}_\mathrm{AP}\triangleq \Big\{\Delta\mathbf{L}\in \mathbb{C}^{N_{\mathrm{PS}}\times {N_{\mathrm{AP}}}}  :\norm{\Delta\mathbf{L}}_\mathrm{F}  \le \upsilon_\mathrm{PS\rightarrow AP}\Big\}, \\
%\hspace*{-6mm}\mathbf{h}_k\hspace*{-2mm}&=&\hspace*{-2mm}\mathbf{\widehat h}_k + \Delta\mathbf{h}_k,\,   \forall k\in\{1,\ldots,K\},\quad  {\bm\Lambda }_k^{\mathrm{AP}} \triangleq \Big\{\Delta\mathbf{h}_k\in \mathbb{C}^{N_{\mathrm{AP}}\times 1}  :\norm{\Delta\mathbf{h}_k}_2  \le \rho_{\mathrm{AP\rightarrow}k}\Big\},\\
%\hspace*{-6mm} \mathbf{f}_k \hspace*{-2mm}&=&\hspace*{-2mm}\mathbf{\widehat f}_k + \Delta\mathbf{f}_k,\,   \forall k\in\{1,\ldots,K\},\mbox{and}\,\,{\bm\Lambda }_k^{\mathrm{PS}} \triangleq  \Big\{\Delta\mathbf{f}_k\in \mathbb{C}^{N_{\mathrm{PS}}\times 1}  :\norm{\Delta\mathbf{f}_k}_2  \le \rho_{\mathrm{PS}\rightarrow k}\Big\}, \forall k, \label{eqn:outdated_CSI-set_end}
\end{eqnarray}
respectively, where $\mathbf{\widehat G}$ and $\mathbf{\widehat E}$ are the  estimates of channel matrices $\mathbf{ G}$ and  $\mathbf{E}$,  respectively,  for resource allocation. Matrices $\Delta\mathbf{G}$ and  $\Delta\mathbf{E}$ represent the  channel uncertainty which captures the joint effects of channel estimation errors and the time varying nature of the associated channels. In particular,  the continuous sets ${\bm\Xi }^{\mathrm{AP}}_\mathrm{E}$ and ${\bm\Xi }^{\mathrm{PS}}_\mathrm{E}$ in \eqref{eqn:outdated_CSI-set1} and \eqref{eqn:outdated_CSI-set2}, respectively,  define the continuous spaces spanned by all possible channel uncertainties with respect to the associated channels. Constants $\upsilon_\mathrm{AP\rightarrow E}$ and $\upsilon_\mathrm{PS\rightarrow E}$, denote the maximum values of the norms of the CSI estimation error matrices $ \Delta\mathbf{G}$ and  $ \Delta\mathbf{E}$, respectively. In practice, the values of $\upsilon_\mathrm{AP\rightarrow E}$ and $\upsilon_\mathrm{PS\rightarrow E}$ depend on the adopted channel estimation algorithms and the coherence times of the associated channels.
\begin{Remark}
Although the eavesdropper may be passive and remain silent, its CSI can be estimated based on the power leakage of the local oscillator
of its  receiver
RF front-end \cite{CN:passive_eave_CSI}.
\end{Remark}

\section{Resource Allocation Problem Formulation}
\label{sec:formulation}
In this section, we first define the system performance metrics and then we formulate the resource allocation algorithm design as an optimization problem.

\subsection{Achievable Data Rate and Secrecy Rate}
Given perfect CSI at the receiver, the achievable data rate of IR $k$ in Phase II is given by
\begin{eqnarray}\label{eqn:user_rate}
R_k &=&  \tau_{\mathrm{II}} \log_2 \big(1+\Gamma_k\big), \, \text{where}\\
\Gamma_k &=& \frac{\mathbf{w}_k^H\mathbf{H}_k\mathbf{w}_k }{\sum_{j \neq k}\mathbf{w}_j\mathbf{H}_k\mathbf{w}_j^H+\underbrace{\Tr(\mathbf{F}_k\mathbf{\Theta})+\Tr(\mathbf{\Psi}\mathbf{H}_k)}_{\text{Interference due to transmitter HWIs }} + \sigma_{d_{\mathrm{IR}_k}}^2+\sigma_n^2}\notag
\end{eqnarray}
is the received signal-to-interference-plus-noise ratio (SINR) at IR $k$. We note that since the pseudo-random artificial noise signals are known to all legitimate transceivers, the impact of the artificial noise signals on the desired signal can be removed at IR $k$ via interference cancellation, i.e., $\Tr(\mathbf{H}_k\mathbf{U})$ and  $\Tr(\mathbf{F}_k\mathbf{Z})$ have been removed\footnote{We note that the interference caused by the transmitter HWIs cannot be removed at IR $k$ as it is an unknown random process.} in \eqref{eqn:user_rate}.

On the other hand, as the computational capability of the eavesdropper is not known, we focus on the worst-case scenario for facilitating secrecy provisioning. In particular, we assume that the eavesdropper is equipped with a noiseless receiver and is able to remove all multiuser interference via successive interference cancellation before attempting to decode the information of IR $k$. As a result, the maximum rate at which the eavesdropper can decode the information intended for IR $k$ is given by
\begin{eqnarray}\label{eqn:eve_rate}
C_{k} &=& \tau_{\mathrm{II}} \log_2 \det \bigg( \mathbf{I}_{N_\mathrm{E}} + \mathbf{Q}^{-1}\mathbf{G}^H\mathbf{w}_k\mathbf{w}_k^H\mathbf{G}\bigg),\, \mbox{where}\\
\mathbf{Q} &=& \mathbf{G}^H(\underbrace{\mathbf{U}+\mathbf{\Psi}}_{\text{Jamming from the AP}})\mathbf{G} +\mathbf{E}^H(\underbrace{\mathbf{Z}+\mathbf{\Theta}}_\text{Jamming from the PS})\mathbf{E}
\end{eqnarray}
is the interference covariance matrix of the eavesdropper.
The achievable secrecy rate between the AP and IR $k$ is given by
\begin{equation}\label{eqn:secrecy_rate}
R_k^{\mathrm{sec}} = \big[ R_k - C_{k} \big] ^{+}.
\end{equation}
As can be seen from \eqref{eqn:eve_rate} and \eqref{eqn:secrecy_rate}, in principle, both the artificial noise signals and the transmitter HWI signals can enhance communication secrecy by degrading the capacity of the channel of the eavesdropper.
\subsection{Total Power Consumption}

In this section, we study the power consumption in both transmission phases. Due to the residual HWIs at the transmitter of the PS, a portion of the transmitted power is wasted during Phase I. More specifically, the total power consumption in Phase I is given by
\begin{eqnarray}\label{eqn:phase-PS-I}
P_{\mathrm{PS-I}} = \rho_{\mathrm{PS}}\bigg( \underbrace{\Tr(\mathbf{V})}_{\text{Power for charging}} + \underbrace{\Tr(\mathbf{\Phi})}_{\text{Power waste}} \bigg)+P_{\mathrm{c}_{\mathrm{PS}}},
%\rho_{\mathrm{PS}} \mathcal{E}\{\norm{\mathbf{v}+\bm{\xi}_{\mathrm{I}}^{(t)}}_2^2\} + P_{\mathrm{c}_{\mathrm{PS}}}\notag\\
%&=&\rho_{\mathrm{PS}}\bigg( \underbrace{\Tr(\mathbf{V}_{\mathrm{I}})}_{\substack{\mathrm{power for}\\\mathrm{charging}}} + \underbrace{\Tr(\mathbf{\Phi})}_{\mathrm{power waste}} \bigg)+P_{\mathrm{c}_{\mathrm{PS}}},
\end{eqnarray}
where $P_{\mathrm{c}_{\mathrm{PS}}}$ accounts for the constant circuit power consumption at the PS. Besides, to capture the power inefficiency of practical power
amplifiers, we introduce a linear multiplicative constant $\rho_{\mathrm{PS}}>1$ for the power radiated by the PS \cite{JR:SWIPT_DAS,JR:QQ_Green_Tutorial}. For example, if $\rho_{\mathrm{PS}}=5$, then for every $1$ Watt of power radiated in
the RF, the PS consumes $5$ Watt of power which leads to a power amplifier efficiency of $20\%$.
In Phase II, the PS transmits artificial noise signals to degrade the channel quality of the eavesdropper and the associated  power consumption is
 \begin{eqnarray}\label{eqn:phase-PS-II}
P_{\mathrm{PS-II}} = \rho_{\mathrm{PS}}\bigg( \underbrace{\Tr(\mathbf{Z})}_{{\text{Power for jamming}}} + \underbrace{\Tr(\mathbf{\Theta})}_{\text{Power waste}} \bigg)+P_{\mathrm{c}_{\mathrm{PS}}}.
\end{eqnarray}

 On the other hand, in Phase II, the AP transmits $K$ independent information signals to the $K$ IRs concurrently. Besides, artificial noise is also emitted by the AP to degrade the channel quality of the eavesdropper. Because of the residual transmitter HWIs, a portion of the power is also wasted  at the AP. Hence, the total power consumption at the AP is given by
\begin{eqnarray}\label{eqn:phase-II-AP}
P_{\mathrm{AP-II}} = \rho_{\mathrm{AP}}\bigg(\underbrace{\sum_{k=1}^{K} \norm{\mathbf{w}_k}^2}_{{\text{Power for information transmission}}} + \underbrace{\Tr(\mathbf{U})}_{{\text{Power for jamming}}} + \underbrace{\Tr(\mathbf{\Psi})}_{\text{Power waste}}\bigg)+P_{\mathrm{c}_{\mathrm{AP}}}.
%\rho_{\mathrm{AP}} \mathcal{E}\{\norm{\sum_{k=1}^{K}\mathbf{w}_k + \mathbf{u}+\bm{\varsigma}^{(t)}}_2^2\} + P_{\mathrm{c}_{\mathrm{AP}}}\\\hspace*{-4mm}&&= \rho_{\mathrm{AP}}\bigg(\underbrace{\sum_{k=1}^{K} \Tr(\mathbf{W}_k)}_{\substack{\mathrm{power for information}\\ \mathrm{transmission}}} + \underbrace{\Tr(\mathbf{V})}_{\substack{\mathrm{power for}\\\mathrm{jamming}}} + \underbrace{\Tr(\mathbf{\Psi})}_{\mathrm{power waste}}\bigg)+P_{\mathrm{c}_{\mathrm{AP}}}.\notag
\end{eqnarray}
Here, $P_{\mathrm{c}_{\mathrm{AP}}}$ is the constant circuit power consumption and $\rho_{\mathrm{AP}} > 1$ denotes the power amplifier inefficiency at the AP.

\subsection{Optimization Problem Formulation}

In the following, we formulate an optimization problem for the minimization of the total power consumption in both transmission phases of the considered WPCN while guaranteeing secure communication in the presence of residual HWIs, imperfect CSI, and a non-linear EH model. The considered optimization problem is given by:
\begin{eqnarray}\label{eqn:power_min_problem_form}
&&\underset{ \substack{\tau_{\mathrm{I}}, \tau_{\mathrm{II}}, \varrho, \mathbf{V}\in \mathbb{H}^{N_{\text{PS}}},\mathbf{Z}\in \mathbb{H}^{N_{\text{PS}}},\\ \mathbf{U}\in \mathbb{H}^{N_{\text{AP}}}, \mathbf{w}_k  }}{\mino}\,\, \tau_{\mathrm{I}}P_{\mathrm{PS-I}}+  \tau_{\mathrm{II}}\Big( P_{\mathrm{AP-II}}+P_{\mathrm{PS-II}}\Big)  \\
\mathrm{s.t.}\,\, &&\mathrm{C1}:\,\,\tau_{\mathrm{II}} \log_2 \big(1 + \Gamma_k\big)\geq R_{\mathrm{req}_k}, \forall k,\notag\\
&&\mathrm{C2}:\,\max_{\Delta\mathbf{G}\in {\bm\Xi }^{\mathrm{AP}}_\mathrm{E}, \Delta\mathbf{ E}\in {\bm\Xi }^{\mathrm{PS}}_\mathrm{E}}\,\tau_{\mathrm{II}}\log_2 \det \bigg( \mathbf{I}_{N_\mathrm{E}} + \mathbf{Q}^{-1}\mathbf{G}^H\mathbf{w}_k\mathbf{w}_k^H\mathbf{G}\bigg) \leq R_{\mathrm{tol}}, \, \forall k,\notag\\
&&\mathrm{C3}:\,\,\tau_{\mathrm{I}} +\tau_{\mathrm{II}} \leq T_{\max}, \notag\\
&&\mathrm{C4}:\,\, \tau_{\mathrm{II}} \bigg(P_{\mathrm{c}_{\mathrm{AP}}}\hspace*{-0.8mm}+\hspace*{-0.8mm} \rho_{\mathrm{AP}} \bigg(\sum_{k=1}^{K}\norm{\mathbf{w}_k}^2 \hspace*{-0.8mm}+ \hspace*{-0.8mm} \Tr(\mathbf{U}) + \Tr (\mathbf{\Psi})\bigg) \bigg) \notag \leq \tau_{\mathrm{I}}\Xi_{\mathrm{tot}}(\varrho) + E_{\mathrm{res}} ,\notag\\
&&\mathrm{C5}:\,\, \varrho \leq \, \Tr(\mathbf{L}^H(\mathbf{V}+\mathbf{\Phi})\mathbf{L}),\,\,\,\,\,\,\quad\quad\quad\quad\quad\quad\mathrm{C6}:\,\, \tau_{\mathrm{I}},\tau_{\mathrm{II}}\geq 0,\notag\\
&&\mathrm{C7}:\,\,\Tr(\mathbf{V})+\Tr(\mathbf{\Phi}) \leq \,P_{\max}^{\mathrm{PS}},\,\,\,\,\quad\quad\quad\quad\quad\quad\mathrm{C8}:\,\, \Tr(\mathbf{Z})  + \Tr(\mathbf{\Theta})\leq P_{\max}^{\mathrm{PS}},\notag\\
&&\mathrm{C9}:\,\, \sum_{k=1}^{K} \norm{\mathbf{w}_k}^2 + \Tr(\mathbf{U}) + \Tr(\mathbf{\Psi})\leq P_{\max}^{\mathrm{AP}}, \,\quad\mathrm{C10}:\,\,  \mathbf{Z},\mathbf{V},\mathbf{U}\succeq \zero. \notag
\end{eqnarray}The objective function in \eqref{eqn:power_min_problem_form} takes into account the total power consumption in Phases I and II at the PS and the AP\footnote{The power consumptions are optimized in both phases, even though the AP is wirelessly charged by the PS only in Phase I. In fact, besides the energy harvested in Phase I, the AP can also use  residual energy, $E_{\mathrm{res}}$, remaining from previous transmission phases for transmission in the current Phase II, cf. constraint C4. Hence, the power consumption of Phase II should also be minimized as well.}, cf. \eqref{eqn:phase-PS-I}--\eqref{eqn:phase-II-AP}. Constraint $\mathrm{C1}$ is imposed such that  the achievable data rate of IR $k$ in Phase II satisfies a minimum required data rate $R_{\mathrm{req}_k}$. On the other hand, taking into account the impact of CSI imperfectness, i.e., sets $\mathbf{\Xi}_{\mathrm{E}}^{\mathrm{AP}}$ and $\mathbf{\Xi}_{\mathrm{E}}^{\mathrm{PS}}$,  constant $R_{\mathrm{tol}}$ in $\mathrm{C2}$ limits the maximum tolerable capacity achieved by Eve in attempting to decode the message of IR $k$. In practice, $R_{\mathrm{req}_k} \gg R_{\mathrm{tol}} > 0$ is set by the system operator to ensure secure communication\footnote{We note that the solution of \eqref{eqn:power_min_problem_form} guarantees a minimum secrecy rate of $R_k^{\mathrm{sec}} =\Big[ R_{\mathrm{req}_k} - R_{\mathrm{tol}}\Big]^+$ for IR $k$ \cite{JR:Xiaoming_magazine_SWIPT}.}. $T_{\max}$ in constraint $\mathrm{C3}$ specifies the total time available for both phases. $\mathrm{C4}$ is a constraint on the overall energy consumption at the AP during Phase II. The total available energy at the AP comprises the energy harvested from the dedicated energy signal transmitted by the PS in Phase I and a constant energy $E_{\mathrm{res}} \geq 0$. In practice,  $E_{\mathrm{res}}$ may represent the residual energy at the AP from previous transmissions or energy obtained from other sources. Furthermore, $\varrho$ in $\mathrm{C5}$  is an auxiliary optimization variable which represents the received RF power at the AP. In particular, $\mathrm{C5}$ ensures that $\varrho$ is always smaller or equal to the minimum harvested power. $\mathrm{C6}$ is a non-negativity constraint on the durations of Phase I and Phase II, respectively. $P_{\max}^{\mathrm{PS}}$ in constraints $\mathrm{C7}$ and $\mathrm{C8}$ limit the maximum transmit power of the PS in Phase I and Phase II, respectively.  Similarly, $P_{\max}^{\mathrm{AP}}$ in constraint $\mathrm{C9}$ specifics the maximum transmit power allowance of the AP in Phase II. Constraint $\mathrm{C10}$, $\mathbf{V}\in \mathbb{H}^{N_{\text{PS}}},\mathbf{Z}\in \mathbb{H}^{N_{\text{PS}}}$, and $ \mathbf{U}\in \mathbb{H}^{N_{\text{AP}}}$ constrain matrices $\mathbf{V},\mathbf{Z}$, and $\mathbf{U}$ to be positive semidefinite Hermitian matrices such that they are valid covariance matrices.

\section{Resource Allocation Algorithm Design}
\label{sec:solution}
The resource allocation problem in \eqref{eqn:power_min_problem_form} is a non-convex optimization problem. In fact, the right-hand side of constraint $\mathrm{C4}$ is a quasi-concave function with respect to $\tau_{\mathrm{I}}$ and $\varrho$. Besides, the optimization variables are coupled in the objective function and in constraint $\mathrm{C4}$. Also, constraint $\mathrm{C2}$ involves infinitely many possibilities due to the uncertainties of the channel estimation errors. Furthermore, the log-det function in constraint $\mathrm{C2}$ is generally intractable.   In this section, we first study the design of a globally optimal resource allocation scheme. The performance of this scheme serves as a performance upper bound for any suboptimal scheme. Then, we derive  a computationally efficient suboptimal resource allocation algorithm.

 \subsection{Optimal Resource Allocation}
To obtain a globally optimal resource allocation scheme, we first perform the following transformation steps. In particular, we introduce auxiliary optimization matrices, $\mathbf{B}_{\mathrm{PS-I}} \in \mathbb{C}^{N_{\mathrm{PS}} \times N_{\mathrm{PS}}}$,  $\mathbf{B}_{\mathrm{PS-II}} \in \mathbb{C}^{N_{\mathrm{PS}} \times N_{\mathrm{PS}}}$, and $\mathbf{B}_{\mathrm{\mathrm{AP}}}\in \mathbb{C}^{N_{\mathrm{AP}} \times N_{\mathrm{AP}}}$, which account for the HWIs at the PS and the AP, respectively. Additionally, we introduce auxiliary optimization variables, $r_{\mathrm{IR},k}, \forall k$, which represent the distortion noise terms caused by the receiver HWIs at the IRs. Then, we transform problem \eqref{eqn:power_min_problem_form} into the following equivalent rank-constrained SDP optimization problem\footnote{In this paper, equivalent means that the transformed problem and the original problem share the same optimal solution.} with optimization variable set $\mathcal{P} = \{  \varrho, \mathbf{V}\in \mathbb{C}^{N_{\mathrm{PS}}} ,\mathbf{Z}\in \mathbb{C}^{N_{\mathrm{PS}}}, \mathbf{U}\in \mathbb{C}^{N_{\mathrm{AP}}} , \mathbf{B}_{\mathrm{PS-I}},\mathbf{B}_{\mathrm{PS-II}}, \mathbf{B}_{\mathrm{AP}}, r_{\mathrm{IR},k}\}$:
\begin{eqnarray}\label{eqn:power_min_problem_reform1}
&&\underset{\tau_{\mathrm{I}}, \tau_{\mathrm{II}},\mathcal{P}, \mathbf{w}_k}{\mino}\quad \tau_{\mathrm{I}}\widetilde{P}_{\mathrm{PS-I}}+  \tau_{\mathrm{II}}(\widetilde{P}_{\mathrm{AP-II}}+\widetilde{P}_{\mathrm{PS-II}}) \\
\mathrm{s.t.}\,\, &&\widetilde{\mathrm{C1}}:\,\,  \frac{\mathbf{w}_k^H\mathbf{H}_k \mathbf{w}_k}{\Gamma_{\mathrm{req}_k}}\geq \sum_{j \neq k} \mathbf{w}_j^H\mathbf{H}_k\mathbf{w}_j \notag\\
&&+\sum_{n=1}^{N_{\mathrm{AP}}}\Tr(\mathbf{S}_n\mathbf{B}_{\mathrm{AP}}\mathbf{H}_k)+
\sum_{n=1}^{N_{\mathrm{PS}}}\Tr(\mathbf{S}_n\mathbf{B}_{\mathrm{PS-II}}\mathbf{F}_k)
+r_{\mathrm{IR},k}+\sigma_n^2, \forall k, \notag \\
&&\widetilde{\mathrm{C2}}:\,\,\max_{\Delta\mathbf{G}\in {\bm\Xi }^{\mathrm{AP}}_\mathrm{E}, \Delta\mathbf{ E}\in {\bm\Xi }^{\mathrm{PS}}_\mathrm{E}}\frac{\mathbf{G}^H \mathbf{w}_k\mathbf{w}_k^H \mathbf{G}}{\Gamma_{\mathrm{tol}}}\preceq  \mathbf{G}^H(\mathbf{U}+\mathbf{B}_{\mathrm{PS}})\mathbf{G}+\mathbf{E}^H(\mathbf{Z}+\mathbf{B}_{\mathrm{PS-II}})\mathbf{E}, \forall k,\notag\\
&&\widetilde{\mathrm{C4}}:\tau_{\mathrm{II}} \widetilde{P}_{\mathrm{AP-II}} \leq \tau_{\mathrm{I}}\Xi_{\mathrm{tot}}(\varrho) + E_{\mathrm{res}} ,\,\mathrm{{C3}},\,\,\mathrm{C6},\,\,\mathrm{C10},\notag\\
&&\widetilde{\mathrm{C5}}:\,\, \varrho \leq  \Tr(\mathbf{L}^H(\mathbf{V}+\mathbf{B}_{\mathrm{PS-I}})\mathbf{L}), \notag\\
&&\widetilde{\mathrm{C7}}:\,\,\Tr(\mathbf{V})+\sum_{m=1}^{N_{\mathrm{PS}}}\Tr(\mathbf{S}_m\mathbf{B}_{\mathrm{PS-I}}) \leq \,P_{\max}^{\mathrm{PS}},\,\,\,\widetilde{\mathrm{C8}}:\,\, \Tr(\mathbf{Z})  + \sum_{m=1}^{N_{\mathrm{PS}}}\Tr(\mathbf{S}_m\mathbf{B}_{\mathrm{PS-II}})\leq P_{\max}^{\mathrm{PS}},\notag\\
&&\widetilde{\mathrm{C9}}:\,\, \sum_{k=1}^{K} \norm{\mathbf{w}_k}^2 + \Tr(\mathbf{U}) + \sum_{n=1}^{N_{\mathrm{AP}}}\Tr(\mathbf{S}_n\mathbf{B}_{\mathrm{AP}})\leq P_{\max}^{\mathrm{AP}},\,
\mathrm{C11: }\,\, \mathbf{B}_{\mathrm{PS-I}},\mathbf{B}_{\mathrm{PS-II}}, \mathbf{B}_{\mathrm{AP}} \succeq \zero,\notag\\
&&\hspace*{-2mm}\mathrm{C12: }\,\,\eta_n\bigg(\sum_{k=1}^{K}[\mathbf{w}_k \mathbf{w}_k^H]_{n,n} + \hspace*{-0.8mm} [\mathbf{U}]_{n,n}\bigg) \leq {[\mathbf{B}_{\mathrm{AP}}]_{n,n}}, \forall n, \notag\\
&&\hspace*{-2mm}\mathrm{C13: }\,\,\eta_m\bigg([\mathbf{V}]_{m,m}  \bigg) \leq {[\mathbf{B}_{\mathrm{PS-I}}]_{m,m}}, \forall m,\,\mathrm{C14: }\eta_m\bigg([\mathbf{Z}]_{m,m}\bigg) \leq {[\mathbf{B}_{\mathrm{PS-II}}]_{m,m}}, \forall m,\notag\\
&&\hspace*{-2mm}\mathrm{C15: }\,\,\nu\bigg(\Tr(\mathbf{H}_k(\mathbf{U}+\sum_{i=1}^{K}\mathbf{w}_i\mathbf{w}_i ^H )+\mathbf{F}_k\mathbf{Z})\bigg)  \leq r_{\mathrm{IR},k},\forall k,\notag
\end{eqnarray}
where $\Gamma_{\mathrm{req}_k}=2^{R_{\mathrm{req}_k}/\tau_{\mathrm{II}}} - 1, \forall k,$ and $\Gamma_{\mathrm{tol}} = 2^{R_{\mathrm{tol}}/\tau_{\mathrm{II}}}-1$ are the equivalent required and the maximum tolerable SINRs at IR $k$ and the eavesdropper, respectively. The power consumption in Phase I and Phase II in the objective function can be rewritten as
\begin{eqnarray}
\widetilde{P}_{\mathrm{PS-I}} \hspace*{-6mm}&&= \rho_{\mathrm{PS}} \bigg(\Tr(\mathbf{V}) + \sum_{m=1}^{N_{\mathrm{PS}}}\Tr(\mathbf{S}_m\mathbf{B}_{\mathrm{PS-I}})\bigg)+P_{\mathrm{c}_{\mathrm{PS}}},\\
\widetilde{P}_{\mathrm{PS-II}} \hspace*{-6mm}&&= \rho_{\mathrm{PS}}\bigg( \Tr(\mathbf{Z}) + \sum_{m=1}^{N_{\mathrm{PS}}}\Tr(\mathbf{S}_m\mathbf{B}_{\mathrm{PS-II}})\bigg) +P_{\mathrm{c}_{\mathrm{PS}}},\quad
\\
\widetilde{P}_{\mathrm{AP-II}} \hspace*{-6mm}&&=\rho_{\mathrm{AP}} \bigg(\hspace*{-0.7mm} \sum_{k=1}^{K} \Tr(\mathbf{w}_k\mathbf{w}_k^H)+ \Tr(\mathbf{U}) + \sum_{n=1}^{N_{\mathrm{AP}}}\Tr(\mathbf{S}_n\mathbf{B}_{\mathrm{AP}})\hspace*{-0.7mm}\bigg)+P_{\mathrm{c}_{\mathrm{AP}}}.
\end{eqnarray}
Furthermore, constraint ${\mathrm{C2}}$ in \eqref{eqn:power_min_problem_form} is replaced by constraint $\widetilde{\mathrm{C2}}$ in \eqref{eqn:power_min_problem_reform1}. These two constraints are equivalent when $R_{\mathrm{tol}} > 0$ and $\Rank(\mathbf{w}_k\mathbf{w}_k^H)\leq 1$, cf. \cite{JR:MOOP_robust_secure_FD}. Constraints $\widetilde{\mathrm{C4}}$, $\widetilde{\mathrm{C5}}$, and $\widetilde{\mathrm{C7}}-\widetilde{\mathrm{C9}}$ are equivalent to the original constraints ${\mathrm{C4}}$, ${\mathrm{C5}}$, and ${\mathrm{C7}}-{\mathrm{C9}}$,  respectively, as the new convex constraints $\mathrm{C12}-\mathrm{C15}$ are satisfied with equality at the optimal solution.

 Next, we handle the coupling of the optimization variables in the objective function and constraint $\widetilde{\mathrm{C4}}$. When both $\tau_{\mathrm{I}}$ and $\tau_{\mathrm{II}}$ are fixed, we can solve \eqref{eqn:power_min_problem_reform1} for the remaining optimization variables. In fact, for a fixed $\tau_{\mathrm{I}}$, the quasi-concavity of the right-hand side of constraint $\widetilde{\mathrm{C4}}$  is also resolved. Besides,  the right-hand side of constraint $\widetilde{\mathrm{C4}}$  is concave with respect to $\varrho$. As a result, we study the optimal resource allocation by assuming that the optimal values of $\tau_{\mathrm{I}}$ and $\tau_{\mathrm{II}}$ satisfying constraints $\mathrm{C3}$ and $\mathrm{C6}$
  are found by a two-dimensional grid search.

Hence, we recast  the  original  problem as an equivalent rank-constrained SDP  optimization problem.
To this end, we define $\mathbf{W}_k=\mathbf{w}_k\mathbf{w}_k^H$ and rewrite the problem in (\ref{eqn:power_min_problem_reform1}) as
\begin{eqnarray}\label{eqn:power_min_problem_reform2}\notag
\hspace*{-8mm}&&\underset{\substack{\mathcal{P},\mathbf{W}_k\in\mathbb{H}^{N_{\mathrm{AP}}}, \bm\Lambda}}{\mino}\quad \tau_{\mathrm{I}}\widetilde{P}_{\mathrm{PS-I}}+  \tau_{\mathrm{II}}(\widetilde{P}_{\mathrm{AP-II}}+\widetilde{P}_{\mathrm{PS-II}}) \\
\mathrm{s.t.}\,\, \hspace*{-5mm}&&\widetilde{\mathrm{C1}}:\,\,  \frac{\Tr(\mathbf{H}_k \mathbf{W}_k)}{\Gamma_{\mathrm{req}_k}}\geq \sum_{j \neq k} \Tr(\mathbf{H}_k\mathbf{W}_j) \notag\\
\hspace*{-8mm}&&+\sum_{n=1}^{N_{\mathrm{AP}}}\Tr(\mathbf{S}_n\mathbf{B}_{\mathrm{AP}}\mathbf{H}_k)+
\sum_{n=1}^{N_{\mathrm{AP}}}\Tr(\mathbf{S}_n\mathbf{B}_{\mathrm{PS-II}}\mathbf{F}_k)
+r_{\mathrm{IR},k}^2+\sigma_n^2, \forall k, \notag \\
\hspace*{-8mm}&&\widetilde{\mathrm{C2}}:\,\,\max_{\Delta\mathbf{G}\in {\bm\Xi }^{\mathrm{AP}}_\mathrm{E}, \Delta\mathbf{ E}\in {\bm\Xi }^{\mathrm{PS}}_\mathrm{E}}\,\frac{\mathbf{G}^H \mathbf{W}_k\mathbf{G}}{\Gamma_{\mathrm{tol}}}\preceq  \mathbf{G}^H(\mathbf{U}+\mathbf{B}_{\mathrm{AP}})\mathbf{G}+\mathbf{E}^H(\mathbf{Z}^H+\mathbf{B}_{\mathrm{PS-II}})\mathbf{E}, \forall k,\notag\\
\hspace*{-8mm}&&\widetilde{\mathrm{C4}}:\tau_{\mathrm{II}} \widetilde{P}_{\mathrm{AP-II}} \leq \tau_{\mathrm{I}}\Xi_{\mathrm{tot}}(\varrho) + E_{\mathrm{res}} ,
\widetilde{\mathrm{C5}}:\,\, \varrho \leq \Tr(\mathbf{L}^H(\mathbf{V}+\mathbf{B}_{\mathrm{PS-I}})\mathbf{L}),\notag\\
\hspace*{-8mm}&& \widetilde{\mathrm{C7}}, \widetilde{\mathrm{C8}}, {\mathrm{C10}}, {\mathrm{C11}},\,\,\notag\
\\
\hspace*{-8mm}&& \widetilde{\mathrm{C9}:}\,\, \sum_{k=1}^{K} \Tr(\mathbf{W}_k) + \Tr(\mathbf{U}) + \sum_{n=1}^{N_{\mathrm{AP}}}\Tr(\mathbf{S}_n\mathbf{B}_{\mathrm{AP}})\leq P_{\max}^{\mathrm{AP}},\notag\\
\hspace*{-12mm}&&\mathrm{C12a: }\,\,\eta_n\big(a_n\big) \leq {[\mathbf{B}_{\mathrm{AP}}]_{n,n}}, \forall n,\,\,\,\,\quad\, \, \mathrm{C12b: }\,\,\sum_{k=1}^{K}\Tr(\mathbf{S}_n\mathbf{W}_k) + \hspace*{-0.8mm} \Tr(\mathbf{S}_n\mathbf{U})\leq a_n, \forall n, \notag\\
\hspace*{-12mm}&&\mathrm{C13a: }\,\,\eta_m\big(b_m\big) \leq {[\mathbf{B}_{\mathrm{PS-I}}]_{m,m}}, \forall m,\,
\,\mathrm{C13b: }\,\,\Tr(\mathbf{S}_m\mathbf{V})   \leq b_m, \forall m,\notag\\
\hspace*{-12mm}&&\mathrm{C14a: }\,\,\eta_m\big(c_m\big) \leq {[\mathbf{B}_{\mathrm{PS-II}}]_{m,m}}, \forall m,\notag\,
\mathrm{C14b: }\,\,\Tr(\mathbf{S}_m\mathbf{Z})\leq c_m, \forall m,\\
\hspace*{-12mm}&&\mathrm{C15a: }\,\,\nu\big(d_k\big)  \leq r_{\mathrm{IR},k},\forall k,\hspace*{18mm}\notag\mathrm{C15b: }\,\,\Tr(\mathbf{H}_k(\mathbf{U}+\sum_{i=1}^{K}\mathbf{W}_i)+\mathbf{F}_k\mathbf{Z}))  \leq d_k,\forall k,\\
\hspace*{-12mm}&& \mathrm{C16: }\,\,\Rank(\mathbf{W}_k)\leq 1,\forall k,\hspace*{18mm}\mathrm{C17: }\,\, \mathbf{W}_k \succeq \zero,\forall k,
 \end{eqnarray}
  where $\bm{\Lambda}=\{a_n,b_m,c_d,d_k,\forall k\}$ is a set of auxiliary optimization variables. We note that the new sets of constraint pairs $\{\mathrm{C12a },\mathrm{C12b }\}$, $\{\mathrm{C13a },\mathrm{C13b }\}$, $\{\mathrm{C14a },\mathrm{C14b }\}$, and $\{\mathrm{C15a },\mathrm{C15b }\}$ are equivalent to the original constraints $\mathrm{C12 }$,  $\mathrm{C13 }$, $\mathrm{C14 }$, and  $\mathrm{C15}$, respectively,
   as the new constraint pairs are satisfied with equality for the optimal solution. Besides,  constraints $\mathrm{C16}$ and $\mathrm{C17}$  are imposed to guarantee that $\mathbf{W}_k=\mathbf{w}_k\mathbf{w}_k^H$  holds for the optimal solution.

Next, we handle the infinitely many possibilities in $\widetilde{\mathrm{C2}}$. First, by introducing an auxiliary optimization matrix $\mathbf{N}\in \mathbb{C}^{N_{\mathrm{EV}}\times N_{\mathrm{EV}}}$, constraint $\widetilde{\mathrm{C2}}$  can be equivalently written as:
\begin{eqnarray}
&&\widetilde{\mathrm{C2a}}:\,\,\max_{\Delta\mathbf{G}\in {\bm\Xi }^{\mathrm{AP}}_\mathrm{E}}\,\frac{\mathbf{G}^H \mathbf{W}_k \mathbf{G}}{\Gamma_{\mathrm{tol}}}\preceq  \mathbf{G}^H(\mathbf{U}+\mathbf{B}_{\mathrm{AP}})\mathbf{G}+\mathbf{N}, \forall k,\\
&&\widetilde{\mathrm{C2b}}:\,\,\min_{ \Delta\mathbf{ E}\in {\bm\Xi }^{\mathrm{PS}}_\mathrm{E}}\,\mathbf{N}\succeq  \mathbf{E}^H(\mathbf{Z}+\mathbf{B}_{\mathrm{PS-II}})\mathbf{E}.
\end{eqnarray}
Then, we introduce a lemma to overcome the infinitely many inequalities in $\widetilde{\mathrm{C2a}}$ and $\widetilde{\mathrm{C2b}}$.
\begin{Lem}\label{lem:extended_S_procedure}[Robust Quadratic Matrix Inequalities \cite{JR:extended_S_procedure}] Let a quadratic matrix function $f(\mathbf{X})$ be defined as
\label{Lemma:S_matrix}
\begin{eqnarray}
f(\mathbf{X})=\mathbf{X}^H\mathbf{A}\mathbf{X}+\mathbf{X}^H\mathbf{B}+\mathbf{B}^H\mathbf{X} +\mathbf{C},
\end{eqnarray}
 where $\mathbf{X},\mathbf{A}, \mathbf{B}$, and $\mathbf{C}$ are arbitrary matrices with appropriate dimensions.  Then, the following two statements are equivalent: \begin{eqnarray}\label{eqm:extended_S_lemma1}
f(\mathbf{X})\succeq \mathbf{0},\forall \mathbf{X}\in\Big\{\mathbf{X}\mid \Tr(\mathbf{D}\mathbf{X}\mathbf{X}^H)\le 1\Big\}\Longleftrightarrow\label{eqm:extended_S_lemma2}\begin{bmatrix}
       \mathbf{C} & \mathbf{B}^H          \\
       \mathbf{B} & \mathbf{A}           \\
           \end{bmatrix} -t\begin{bmatrix}
       \mathbf{I} & \mathbf{0}          \\
       \mathbf{0} & -\mathbf{D}           \\
           \end{bmatrix}          \succeq \mathbf{0},  \,\,\mbox{if } \exists t\ge 0,
\end{eqnarray}
for matrix $\mathbf{D}\succeq \zero$ and $t$ is an auxiliary constant.
\end{Lem}

Then, constraint $\widetilde{\mathrm{C2a}}$ can be equivalently transformed into:
\begin{eqnarray}\label{eqn:LMI_C5}
          \widetilde{\mathrm{C2a}}&:&\mathbf{S}_{\mathrm{C{2a}}_{k}}
          \Big({\mathbf{B}}_{\mathrm{AP}},\mathbf{U},\mathbf{W}_k,\mathbf{N},t_k\Big)=
          \begin{bmatrix}
       \mathbf{N}-t_k\mathbf{I}_{N_{\mathrm{EV}}}&  \zero        \\
          \zero  & \hspace*{-1mm} \frac{t_k\mathbf{I}_{\mathrm{AP}}}{\upsilon_\mathrm{AP\rightarrow E}^2}  \\
           \end{bmatrix}+           \mathbf{R}_{\mathbf{\widehat G}}^H\mathbf{M}_k\mathbf{R}_{\mathbf{\widehat G}}       \succeq \mathbf{0}, \forall k, \notag\\
           \mbox{where}\quad \mathbf{M}_k&=&\mathbf{U}+\mathbf{B}_{\mathrm{AP}}-\frac{\mathbf{W}_k}{\Gamma_{\mathrm{tol}}}\quad \mbox{and}\quad \mathbf{R}_{\mathbf{\widehat G}}=[ \mathbf{\widehat G}\,\,\,\, \mathbf{I}_{\mathrm{AP}}].
\end{eqnarray}

Similarly, we can transform constraint $\widetilde{\mathrm{C2b}}$ into  its equivalent form:
\begin{eqnarray}\label{eqn:LMI_C2b}
          \widetilde{\mathrm{C2b}}&:&\mathbf{S}_{\mathrm{C2b}}
          \Big({\mathbf{B}}_{\mathrm{PS-II}},\mathbf{Z},\mathbf{N},\gamma\Big)=
          \begin{bmatrix}
       -\mathbf{N}-\gamma\mathbf{I}_{\mathrm{PS}}&  \zero        \\
           \zero & \hspace*{-1mm} \frac{\gamma \mathbf{I}_{\mathrm{PS}}}{\upsilon_\mathrm{PS\rightarrow E}^2}  \\
           \end{bmatrix} +          \mathbf{R}_{\mathbf{\widehat E}}^H\mathbf{K}\mathbf{R}_{\mathbf{\widehat E}}       \succeq \mathbf{0}, \forall k, \notag\\
           \mbox{where}\quad \mathbf{K}&=&\mathbf{Z}+\mathbf{B}_{\mathrm{PS-II}}\quad \mbox{and}\quad \mathbf{R}_{\mathbf{\widehat E}}=[ \mathbf{\widehat E}\,\,\,\, \mathbf{I}_{\mathrm{PS}}].
\end{eqnarray}

 Next, we relax the non-convex constraint in $\mathrm{C16}$ by removing it from the problem formulation. Therefore, for a given $\tau_{\mathrm{I}}$ and  $\tau_{\mathrm{II}}$, the equivalent SDP relaxed formulation of \eqref{eqn:power_min_problem_reform1} is given by:
\begin{eqnarray}\label{eqn:power_min_problem_reform2}
&&\underset{\mathcal{P},\mathbf{W}_k\in\mathbb{H}^{N_{\mathrm{AP}}}, \bm\Lambda,\mathbf{N}\in \mathbb{H}^{N_{\mathrm{EV}}},t_k,\gamma}{\mino}\, \quad \tau_{\mathrm{I}}\widetilde{P}_{\mathrm{PS-I}}+  \tau_{\mathrm{II}}(\widetilde{P}_{\mathrm{AP-II}}+\widetilde{P}_{\mathrm{PS-II}})\\
\mathrm{s.t.}\,\, &&\widetilde{\mathrm{C1}}, \, \widetilde{\mathrm{C2a}}, \,\widetilde{\mathrm{C2b}}, \widetilde{\mathrm{C4}}, \, \widetilde{\mathrm{C5}},  \,\widetilde{\mathrm{C7}}-\widetilde{\mathrm{C9}}, \, \mathrm{C10, C11}, \mathrm{C12a },\mathrm{C12b }, \mathrm{C13a },\mathrm{C13b },\mathrm{C14a },\mathrm{C14b },\notag\\\notag
&&\mathrm{C15a },\mathrm{C15b },\mathrm{C18:}t_k,\gamma\geq 0,\notag
\end{eqnarray}
where constraint $\mathrm{C18}$ is due to the use of Lemma  \ref{lem:extended_S_procedure}  for handling the infinitely  many constraints associated with the channel estimation errors.

The optimization problem in \eqref{eqn:power_min_problem_reform2} is a standard convex optimization problem and can be solved efficiently by numerical convex program solvers such as CVX \cite{website:CVX}. However, by solving \eqref{eqn:power_min_problem_reform2} numerically, in general,  there is no guarantee that the optimal solution of \eqref{eqn:power_min_problem_reform2} satisfies $\mathrm{C16}$ of the original problem formulation in \eqref{eqn:power_min_problem_form}, i.e., $\Rank(\mathbf{W}_k)\leq 1$. Hence, in the following, we study the structure of the solution of the SDP relaxed problem in \eqref{eqn:power_min_problem_reform2}.
\begin{Thm}\label{thm:equality}Suppose that the optimization problem in \eqref{eqn:power_min_problem_reform2} is feasible. Then, for $\Gamma_{\mathrm{req}_k} > 0$ and $\Gamma_{\mathrm{tol}}>0$, the rank-one constraint relaxation of $\mathbf{W}_k$, of the SDP relaxed optimization problem in \eqref{eqn:power_min_problem_reform2} is tight, i.e., $\Rank(\mathbf{W}_k) = 1, \forall k$.   Besides, $\Rank(\mathbf{U}) \leq 1$, $\Rank(\mathbf{Z}) \leq 1$, $\Rank(\mathbf{V}) \leq 1$.
\end{Thm}

\emph{Proof: }Please refer to the Appendix.

Theorem \ref{thm:equality} states that the globally optimal solution of \eqref{eqn:power_min_problem_reform2} can be obtained by information beamforming for each IR, despite the HWIs at the transceivers. Moreover, beamforming is also optimal for wireless charging and jamming in Phase I and Phase II, respectively, even when the non-linearity of the EH circuits, the residual HWIs, and the imperfect CSI are taken into account. In summary, we first discretize the continuous optimization variables $\tau_{\mathrm{I}},\tau_{\mathrm{II}}$. Then,  we solve \eqref{eqn:power_min_problem_reform2} for each pair of $\tau_{\mathrm{I}},\tau_{\mathrm{II}}$ satisfying $\tau_{\mathrm{I}}+\tau_{\mathrm{II}} \in [0, T_{\mathrm{max}}]$. Finally, we obtain the globally optimal solution\footnote{Note that the optimality of this method depends on the resolution of the discretization. } of \eqref{eqn:power_min_problem_form} by employing a two-dimensional search over all combinations of  $\tau_{\mathrm{I}},\tau_{\mathrm{II}}$ to find the minimum objective value.

\begin{table}[t]\vspace*{-0.3cm}\caption{Iterative Resource Allocation Algorithm}\label{table:algorithm}
\vspace*{-0.6cm}
\renewcommand\thealgorithm{}
\begin{algorithm} [H]                    % enter the algorithm environment
\caption{Alternating Optimization}          % give the algorithm a caption
\label{alg1}                           % and a label for \ref{} commands later in the document
\begin{algorithmic} [1]
      % enter the algorithmic environment
\STATE Initialize the maximum number of iterations $L_{\max}$ and convergence error tolerance $\psi\rightarrow 0$
\STATE Set iteration index $l=0$ and initialize $\{\mathcal{P},\mathbf{W}_k, \bm\Lambda,\mathbf{N},t_k,\gamma\}$

\REPEAT [Loop]
%%%%%%%%%%%%%%%%%%%%

\STATE Solve  (\ref{eqn:AO_tau})  for  $\tau_\mathrm{I}$ and $\tau_\mathrm{II}$ for given $\{\mathcal{P}',\mathbf{W}_k', \bm\Lambda',\mathbf{N}',t_k',\gamma'\}=\{\mathcal{P},\mathbf{W}_k, \bm\Lambda,\mathbf{N},t_k,\gamma\}$  which leads to  intermediate time allocation variables $\tau_\mathrm{I}'$ and $\tau_\mathrm{II}'$

\STATE Solve  (\ref{eqn:power_min_problem_reform2})  for  $\{\mathcal{P},\mathbf{W}_k, \bm\Lambda,\mathbf{N},t_k,\gamma\}$ for given  $\tau_\mathrm{I}'$ and $\tau_\mathrm{II}'$ via SDP relaxation and obtain the intermediate solution  $\{\mathcal{P}',\mathbf{W}_k', \bm\Lambda',\mathbf{N}',t_k',\gamma'\}$
%%%%%%%%%%%%%%%%%%%%
\IF{$\abs{\tau_\mathrm{I}'-\tau_\mathrm{I}(l-1)}\le \psi$ and $\abs{\tau_\mathrm{II}'-\tau_\mathrm{II}(l-1)}\le \psi$ }
\STATE
Convergence = \TRUE \RETURN
$\{\mathcal{P}',\mathbf{W}_k', \bm\Lambda',\mathbf{N}',t_k',\gamma',\tau_\mathrm{I}',\tau_\mathrm{II}'\}$
\ELSE
\STATE $\tau_\mathrm{I}(l)=\tau_\mathrm{I}'$, $\tau_\mathrm{II}(l)=\tau_\mathrm{II}'$, $l=l+1$
 \ENDIF
\UNTIL{ $l=L_{\max}$}

\end{algorithmic}
\end{algorithm}
\normalsize\vspace*{-15mm}
\end{table}
%%%%%%%%%%%%%%%%%%%%%%%%%%%%%%%%%%%%%%%%%%%%%%%%%%%%%%%%%%%%%%%%%%%%%%%%%%%%%%%
\subsection{Suboptimal Solution}
Although the method proposed in the last section achieves the globally optimal solution of \eqref{eqn:power_min_problem_form}, it requires a two-dimensional search with respect to optimization variables $\tau_{\mathrm{I}}, \tau_{\mathrm{II}}$ and the number of SDPs to be solved increase quadratically with the resolution of the search grid. To reduce the computational complexity, we propose in the following a suboptimal resource allocation algorithm. In fact, \eqref{eqn:power_min_problem_form} is jointly convex with respect to $\tau_{\mathrm{I}}$ and $\tau_{\mathrm{II}}$ when the other optimization variables are fixed. As a result, an iterative
alternating optimization method  \cite{JR:AO} is proposed to obtain a locally optimal
solution  of  \eqref{eqn:power_min_problem_form}  and  the  algorithm  is  summarized  in  Table \ref{table:algorithm}.
The algorithm is implemented by a repeated loop.  In line 2,  we first set the iteration index $l$ to zero
and initialize the resource allocation policy. Variables $\tau_\mathrm{I}(l)$ and $\tau_\mathrm{II}(l)$ denote the time allocation policy in the $l$-th iteration. Then, in each iteration, for a given intermediate beamforming policy $\{\mathcal{P}',\mathbf{W}_k', \bm\Lambda',\mathbf{N}',t_k',\gamma'\}$, we solve
\begin{eqnarray}\label{eqn:AO_tau}\notag
&&\underset{\{\tau_{\mathrm{I}},\tau_{\mathrm{II}}\}}{\mino}\, \quad \tau_{\mathrm{I}}\widetilde{P}_{\mathrm{PS-I}}+  \tau_{\mathrm{II}}(\widetilde{P}_{\mathrm{AP-II}}+\widetilde{P}_{\mathrm{PS-II}})\\
\mathrm{s.t.}\,\, &&\widetilde{\mathrm{C1}}, \, \widetilde{\mathrm{C2a}}, \,\widetilde{\mathrm{C2b}}, {\mathrm{C3}}, \widetilde{\mathrm{C4}}, {\mathrm{C6}},
\end{eqnarray}
 c.f., line $4$ of Table I.  Since \eqref{eqn:AO_tau} is a linear programming (LP) problem, we can solve \eqref{eqn:AO_tau} via the simplex method or any standard numerical solver for solving LPs \cite{book:convex}.   The  time allocation obtained from \eqref{eqn:AO_tau}, i.e., $\tau_{\mathrm{I}}'$ and $\tau_{\mathrm{II}}'$, is used as an input to  (\ref{eqn:power_min_problem_reform2})  for solving for $\{\mathcal{P}',\mathbf{W}_k', \bm\Lambda',\mathbf{N}',t_k',\gamma'\}$ via SDP relaxation. Then,  we repeat the procedure iteratively until
 the maximum number of iterations is reached or convergence is
achieved. We note convergence to a locally optimal solution of (\ref{eqn:power_min_problem_form}) is guaranteed for a sufficiently large number of iterations \cite{JR:AO}. Besides, the proposed suboptimal algorithm has a polynomial time computational complexity and does not require any form of exhaustive search.

\begin{table}[t]
\caption{Simulation Parameters.} \label{table:parameters}
\centering
\vspace*{-2mm}
\begin{tabular}{ | L | l | } \hline
      Carrier center frequency                           &  915 MHz\\ \hline
      Bandwidth                                          &  $200$ kHz \\ \hline
      Path loss exponent: PS$\rightarrow$AP, AP$\rightarrow$IRs,  PS$\rightarrow$Eve, AP$\rightarrow$Eve								 &  $2$, $3.6$, $3.6$, $3.6$ \\ \hline
            PS to AP fading distribution                                      & Rician, Rician factor $3$ dB \\ \hline
          AP to IRs, Eve fading distribution                                      & Rayleigh \\ \hline
      PS and AP antenna gain                       &  $10$ dBi and $8$ dBi \\ \hline
      Noise power                                    & $\sigma_n^{\textnormal{\tiny \textsc{2}}}=\sigma_{{{\textnormal{\tiny \textsc{ir}}}}}^{\textnormal{\tiny \textsc{2}}} = \sigma_{{{\textnormal{\tiny \textsc{e}}}}}^{\textnormal{\tiny \textsc{2}}} =  -77$ dBm \\ \hline
      Power amplifier efficiency 						 &  ${1/\rho_{{{\textnormal{\tiny \textsc{ps}}}}} = 1/\rho_{{{\textnormal{\tiny \textsc{ap}}}}} = 30\, \%}$ \\ \hline
      Circuit power consumption						 &  $P_{\textsc{c}_{{\textnormal{\tiny \textsc{ps}}}}}$	$ = P_{\textsc{c}_{{\textnormal{\tiny \textsc{ap}}}}} = 50\, \mu$W \\ \hline
      Non-linear EH model  parameters   &  $M = 24$ mW, $a=150$,   $b = 0.0014$ \cite{CN:EH_measurement_2}  \\ \hline
      Distance PS-to-AP,  PS-to-Eve, AP-to-Eve  &  $10$ m, $40$ m, $30$ m  \\
     \hline

     Maximum transmit power at PS and AP &$P_{\max}^{\mathrm{PS}}=P_{\max}^{\mathrm{AP}}=30$ dBm\\
      \hline
\end{tabular}\vspace*{-6mm}
\end{table}
\section{Simulation}
In this section, we evaluate the performance of the proposed optimal and suboptimal resource allocation schemes for the considered WPCN architecture. The relevant simulation parameters are provided in Table \ref{table:parameters}.  For conducting the two-dimensional search for the optimal resource allocation, we quantize the possible ranges of $\tau_{\text{I}},\tau_{\text{II}}$, $0.0001 \leq \tau_{\text{I}},\tau_{\text{II}} \leq T_{\text{max}}$, into  $20\times 20$  equally  spaced  intervals, and for simplicity, we normalize the duration of the communication slot to $T_{\text{max}} = 1$.  Unless specified otherwise, we assume for the transmitter HWI parameters $k_{\mathrm{1}}=2.258\times 10^{5}$ and  $k_{\mathrm{2}}=7.687$, and for the receiver distortion parameter $k_{\text{3}}=0$.  The IRs are randomly distributed at a distance of $50$ meters around the AP and the data rate requirements of all IRs are equal, i.e., $R_{\text{req}_k} = R_{\text{req}}$ bit/s, $\forall k$, while the maximum tolerable rate of Eve is $R_{\text{tol}} = 0.1 $ bit/s. Besides, we assume $N_{\text{PS}} = 6$, $N_{\text{AP}} = 6$, and $N_{\text{EV}} = 2$ antennas at the PS, AP, and Eve, respectively, unless specified otherwise. For calculating the system power consumption, to avoid counting the same power twice, we consider only the power consumption of Phase I and the power consumed from  $E_{\mathrm{res}}$ (if any)  in Phase II. In the sequel,  we define the normalized maximum  channel estimation error of the eavesdropper  as  $\sigma_{\mathrm{EVE}}^2=1\%\geq\frac{\upsilon_\mathrm{AP\rightarrow E}^2}{\norm{\mathbf{G}}^2_\mathrm{F}},\frac{\upsilon_\mathrm{PS\rightarrow E}^2}{\norm{\mathbf{E}}^2_\mathrm{F}}$. Besides, the results shown in this section were averaged over $10000$ fading channel  realizations.

\subsection{Convergence of Iterative Suboptimal Algorithm}

\begin{figure}[t]
\centering\vspace*{-4mm}
\includegraphics[width=4.5 in]{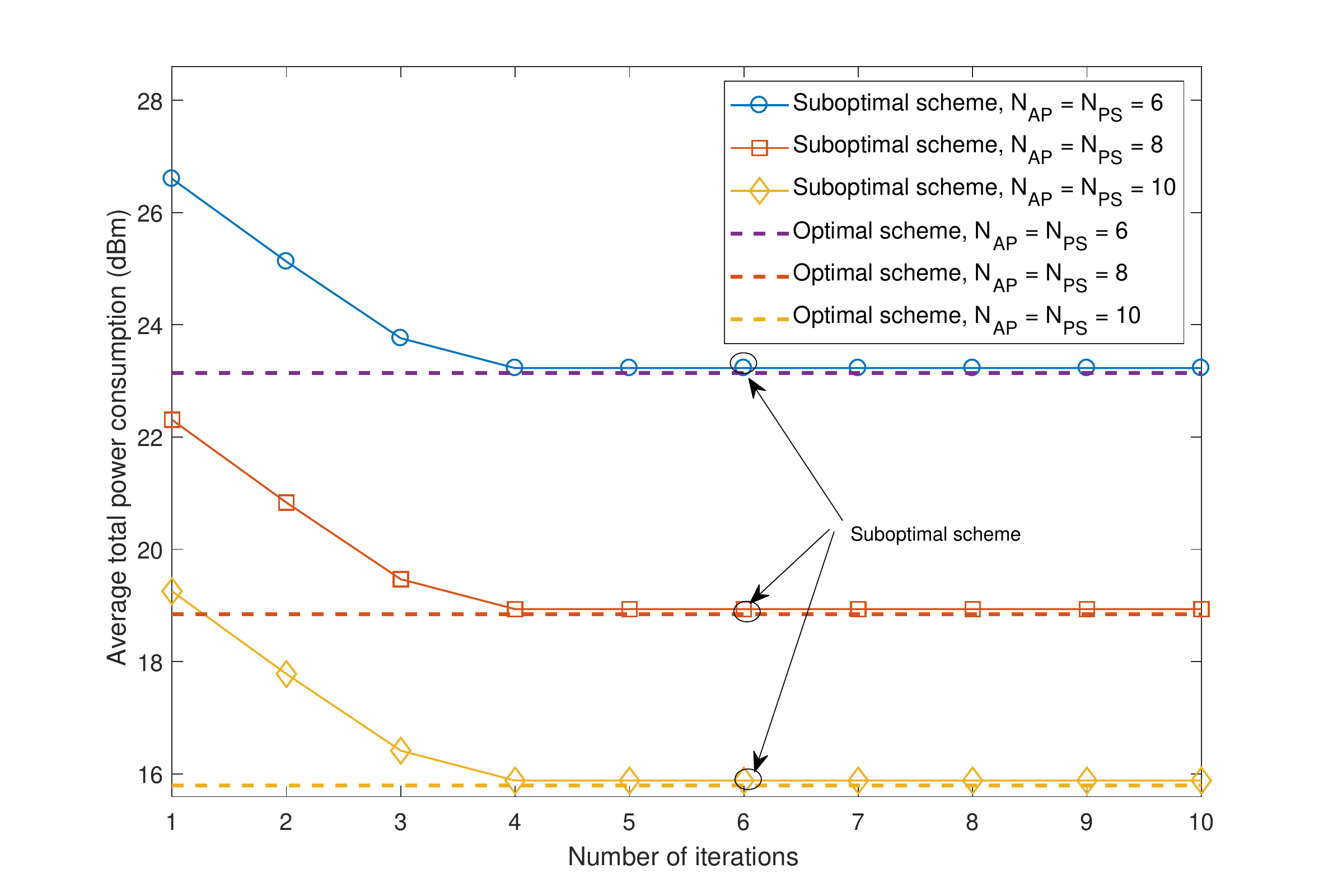}\vspace*{-4mm}
\caption{Average total power consumption (dBm) versus the number of iterations for different numbers of antennas equipped at the PS and the AP.}\vspace*{-6mm}
\label{fig:convergence}
\end{figure}

Figure \ref{fig:convergence} illustrates the convergence behavior of the proposed iterative suboptimal algorithm for different numbers of antennas equipped at the PS and the AP. We set the minimum required data rate per IR to $R_{\text{req}}=8$ bits/s/Hz. As can be observed, the proposed iterative suboptimal
algorithm converges  within $10$
iterations on average for all considered scenarios. In particular, the performance of the suboptimal scheme closely approaches  that of the optimal scheme. On the other hand, the numbers of antennas equipped at the AP and the PS have only a small impact on the speed of convergence.

In the sequel,  for studying the system performance,  we set the number of iterations for the proposed suboptimal algorithm
to $10$.

\subsection{Average Total Transmit Power versus Minimum Required Rate}
\begin{figure}[t]
\centering\vspace*{-4mm}
\includegraphics[width=4.5 in]{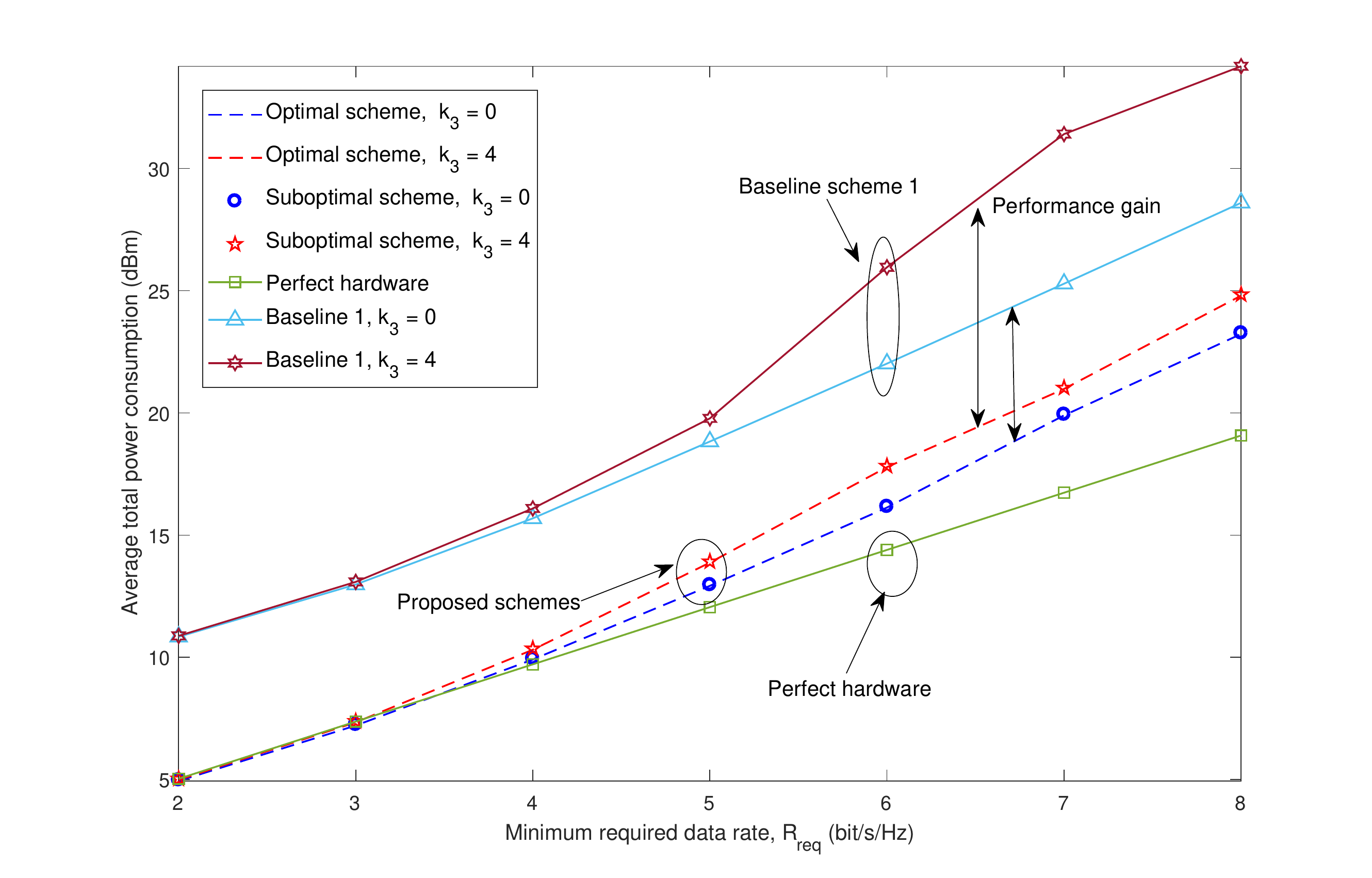}\vspace*{-4mm}
\caption{Average total power consumption (dBm) versus  minimum required data rate per IR, $R_{\mathrm{req}}$, for different values of receiver HWI parameter, $k_3$. The double-sided arrows
indicate the power saving due to the proposed optimization.}
\label{fig:power-vs-R-min}\vspace*{-8mm}
\end{figure}

In Figure \ref{fig:power-vs-R-min}, we show the total average power consumption versus the minimum required data rate per IR,  $R_{\text{req}}$, for different values of receiver HWI parameter, $k_{\text{3}}$. Zero residual energy is assumed, i.e., $E_{\text{res}}=0$. As can be seen from Figure \ref{fig:power-vs-R-min}, the total average power consumption of the proposed optimal and suboptimal schemes increases monotonically with $R_{\text{req}}$.  The reasons behind this are twofold. First, a higher transmit power for the information signals, $\mathbf{w}_ks_k$,  is necessary in order to meet more stringent requirements on the minimum data rate.   Second, more power has to be allocated to the artificial noise,
 $\mathbf{z}$, for neutralizing the increased information leakage due to the
higher power of $\mathbf{w}_ks_k$. Hence, the PS has to increase the transmit power for wireless charging in Phase I to ensure that a sufficient amount of energy is available for wireless information transfer and artificial noise generation in Phase II. Besides, it can be observed that the average total  power consumption increases for increasing receiver HWI parameter, $k_3$.
 In  fact, for  the same amount of received power,  the received SINR deteriorates with an increasing  receiver HWI parameter, $k_3$. Hence,  the AP has to sacrifice  some spatial degrees of freedom used for mitigation of the multiuser interference received at each IR to alleviate the impact of receiver HWI. As a result, it becomes more challenging for the AP to focus
the energy of information signals on the IRs which results in a higher transmit power in Phase I and Phase II.  On the other hand, although only $10$ iterations are used, the proposed suboptimal scheme performs close to the optimal scheme employing the two-dimensional exhaustive search.

For comparison, Figure \ref{fig:power-vs-R-min} also contains the performance of one benchmark scheme and two baseline schemes.
For the benchmark scheme,   we assume that  perfect hardware is available at all transceivers of the considered WPCN, i.e., there are no HWIs, and the corresponding performance serves as an upper bound for the proposed schemes.  For baseline $1$, we set $P_{\max}^{\mathrm{PS}}=P_{\max}^{\mathrm{AP}}=40$ dBm. Then, we solve \eqref{eqn:power_min_problem_form} with the proposed optimal resource allocation algorithm but adopt a fixed isotropic radiation pattern for  $\mathbf{V}$.  We also considered a  baseline  $2$ where resource allocation was performed subject to the same constraint set as in \eqref{eqn:power_min_problem_form}, except that  the residual HWIs at the transceivers were not considered. The performance of baseline  $2$ was then evaluated in the presence of the residual HWIs.  However,   for the adopted simulation
settings, baseline $2$
could not satisfy the QoS requirements in constraints $\mathrm{C1}$ and $\mathrm{C2}$ as the residual HWIs were ignored in the design phase. Therefore, performance results for baseline $2$ are not shown in Figure \ref{fig:power-vs-R-min}. This underlines the importance of taking residual HWIs into account for resource allocation design.   On the other hand, as expected, the  performance gap between the scheme with the perfect hardware and the proposed schemes is slightly enlarged as $R_{\mathrm{req}}$ increases. In fact, the interference caused by the HWIs at transmitter and receiver increases with the transmit power  and the received power, respectively. Hence, a higher data rate requirement, $R_{\mathrm{req}}$, magnifies the impact of the HWIs on system performance. Furthermore,  as can be observed from Figure \ref{fig:power-vs-R-min}, baseline $1$ consumes a significantly higher power compared to the proposed schemes. This is because  baseline  $1$ is less efficient in wireless charging compared to the proposed schemes. In particular, the
PS and the AP cannot fully utilize the available degrees of freedom since the direction of
the transmit energy signal at the PS in Phase I is  fixed. The resulting performance gap reveals the
importance of optimizing all beamforming matrices  for
the minimization of  the total power consumption of the considered WPCN.

%\begin{figure}[t]
% \centering
% \hspace*{-15mm}\begin{minipage}[b]{0.40\linewidth}  \hspace*{-12mm}
%\includegraphics[width=3.5 in]{time_allocation.pdf}\vspace*{-3mm}
%\caption{Average Phase I and Phase II durations versus the maximum transmit power at the BS (dBm).  }\label{fig:time_allocation}
% \end{minipage}\hspace*{1.1cm}
% \begin{minipage}[b]{0.40\linewidth} \hspace*{-0.7cm}
%\includegraphics[width=3.5 in]{power_allocation_vs_req.pdf}\vspace*{-3mm}\hspace*{1.6cm}
%\caption{Average system throughput (Mbits/s) versus the maximum transmit power at the BS (dBm). } \label{fig:power_allocation}
% \end{minipage}
%\end{figure}

\begin{figure}[t]\centering\vspace*{-8mm}
\includegraphics[width=4.5 in]{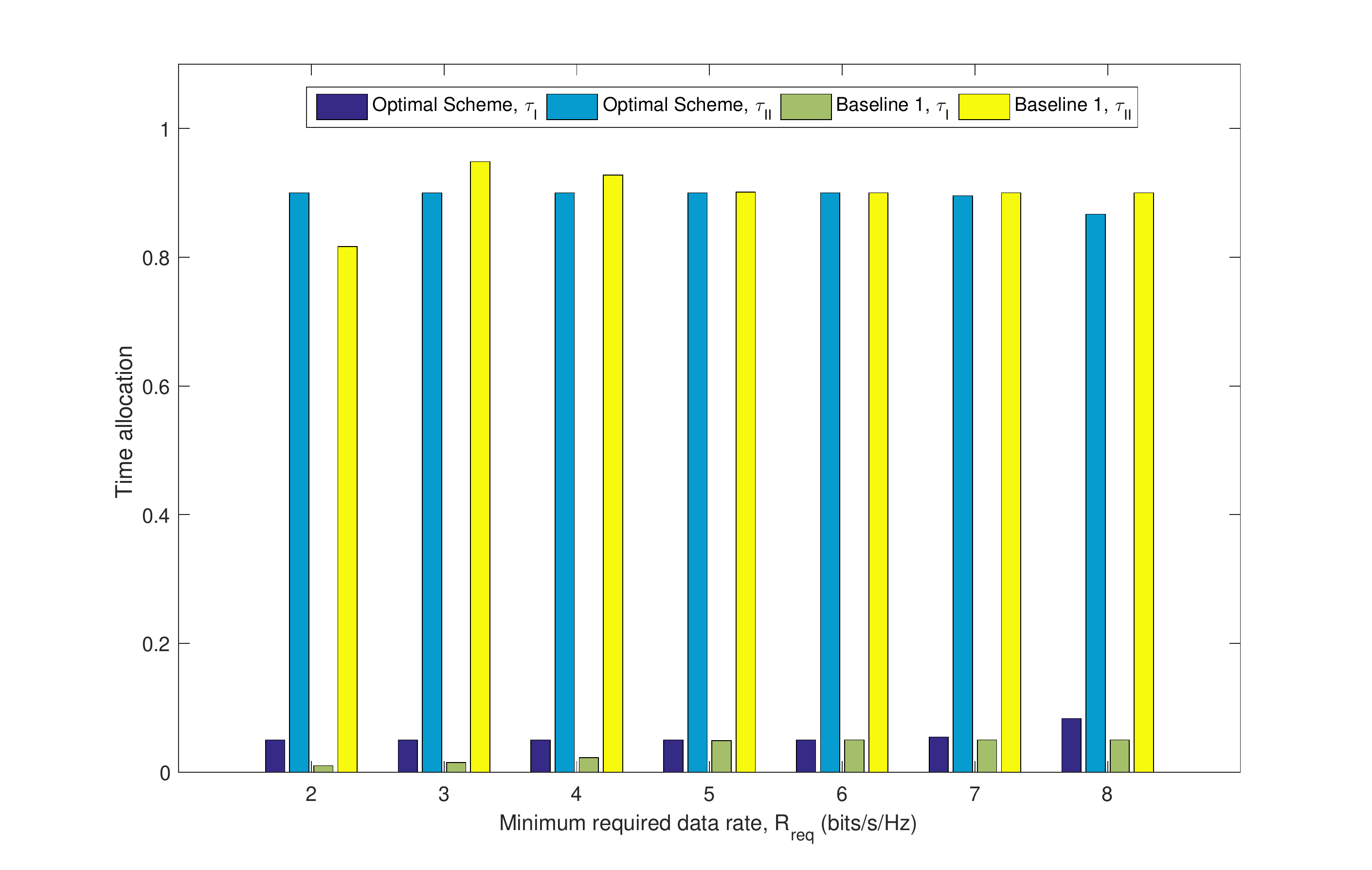}\vspace*{-3mm}
\caption{Average time allocation  versus minimum required data rate per IR, $R_{\mathrm{req}}$, for $k_{\mathrm{3}}=0$.   }\label{fig:time_allocation}
\includegraphics[width=4.5 in]{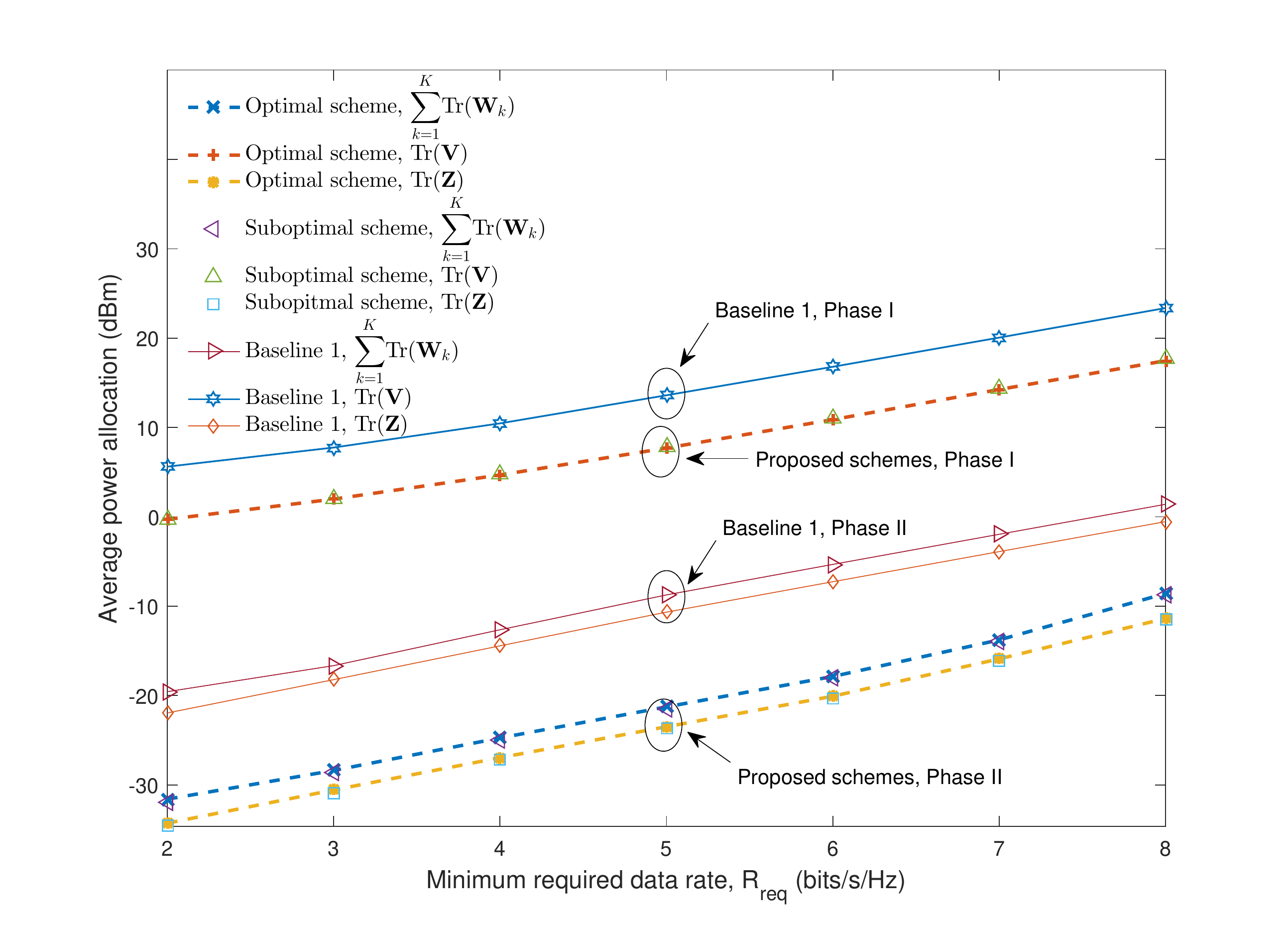}\vspace*{-3mm}
\caption{Average power allocation (dBm) versus minimum required data rate per IR, $R_{\mathrm{req}}$, for $k_{\mathrm{3}}=0$. } \label{fig:power_allocation}\vspace*{-6mm}
\end{figure}

In Figures \ref{fig:time_allocation} and \ref{fig:power_allocation}, we depict the time allocation and power allocation of the proposed schemes and baseline $1$ for the scenario with $k_{\mathrm{3}}=0$ studied in Figure \ref{fig:power-vs-R-min}. In particular, Figure \ref{fig:power_allocation} shows the average transmit powers allocated to the three components\footnote{Since $\Tr(\mathbf{U})=0$ in all the considered cases, it is not shown in Figure \ref{fig:power_allocation}. } of the transmitted signals, i.e., $\sum_{k=1}^K\Tr(\mathbf{W}_k)$, $\Tr(\mathbf{V})$, and $\Tr(\mathbf{Z})$. First, it can be observed from Figure \ref{fig:time_allocation} that the time allocated\footnote{Note that the time allocation for the proposed suboptimal scheme is similar to that of the optimal scheme, and  hence, is omitted for brevity. } to Phase I for wireless charging in the considered WPCN is monotonically increasing with respect to the minimum data rate requirement per IR.  In fact, the system increases both the transmit power and the time duration allocated to the PS in Phase I to enable a more effective wireless charging of the AP as the data rate requirement becomes more stringent. Besides, for all considered values of $R_{\mathrm{req}}$, it can be seen that a small portion of time is allocated for wireless charging while a large portion of time is reserved for wireless information transfer. Recall that the equivalent minimum required SINR at IR $k$ is defined as $\Gamma_{\mathrm{req}_k}=2^{R_{\mathrm{req}_k}/\tau_{\mathrm{II}}} - 1$. Hence, increasing the value of $\tau_{\mathrm{II}}$ can effectively lower the required equivalent SINR so as to reduce the total system power consumption. On the other hand,   as expected, the amounts of power allocated  to the
information signals and the energy/artificial noise signals, $\mathbf{v},\mathbf{z}$, increase  for the proposed schemes as the minimum required data rate per IR $R_{\mathrm{req}}$ increases. In particular,  the power allocated to the energy/artificial noise signals, $\mathbf{v},\mathbf{z}$, increases
as fast as the power allocated to the information signals when $R_{\mathrm{req}}$ increases. This is because for a more stringent data rate requirement, a higher transmit power is needed for information transmission, and thus, the considered WPCN system is more vulnerable to eavesdropping. Hence, also more energy has to be allocated to the energy/artificial noise signals for   more  effective wireless charging and jamming.  Interestingly, $\Tr(\mathbf{U})=0$ holds for all considered values of the required data rate which suggests that generating artificial noise at the AP for jamming is not beneficial. In fact, transmitting artificial noise from the AP for ensuring communication security is less power efficient compared to transmitting the artificial noise directly from the PS.
The reasons for this are twofold. First, the non-linearity of the  energy harvesting circuits limits the maximum amount of harvestable energy for generating a sufficiently strong  jamming signal at the AP. Second, since $E_{\mathrm{res}}=0$, all  energy consumed at the AP has to be harvested first from the RF signals transmitted by the PS. As a result, if the AP utilizes the harvested energy in Phase I to generate a jamming signal in Phase II,  the energy of the jamming signal is subject to the  attenuation in both the PS-to-AP link and the AP-to-Eve link before  reaching the potential eavesdropper.  Such a ``double energy attenuation" severely decreases the efficiency of communication security provisioning, and thus is avoided by the optimal resource allocation scheme.

%\begin{figure}[t]
%\centering
%\includegraphics[width=4 in]{time_allocation_graph_v3aaa.pdf}
%\caption{Average total power consumption (dBm) versus the number of antennas, $N_{\mathrm{AP}},N_{\mathrm{PS}}$  for different amounts of available residual energy $E_{\mathrm{res}}$.}\label{fig:power_vs_antennas}\end{figure}

\begin{figure}[t]\centering\vspace*{-8mm}
\includegraphics[width=4.5 in]{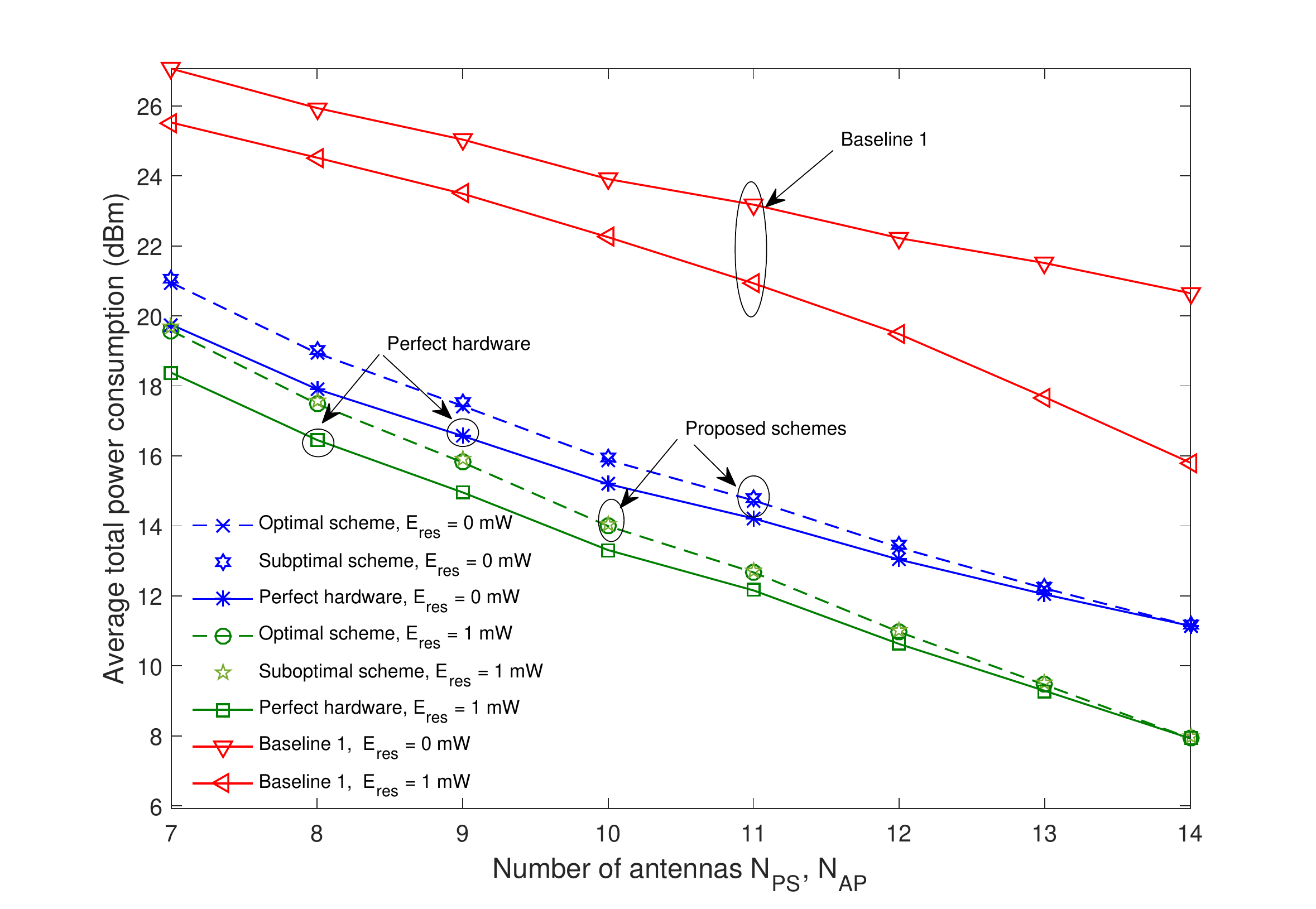}
\caption{Average total power consumption (dBm) versus the numbers of antennas equipped at the PS and the AP, $N_{\mathrm{AP}}$ and $N_{\mathrm{PS}}$, for different amounts of available residual energy $E_{\mathrm{res}}$. }
\label{fig:power_vs_antennas}\vspace*{-5mm}
\end{figure}

In Figure \ref{fig:power_vs_antennas}, we show the average total power consumption versus the number of antennas equipped at the PS and the AP for  different resource allocation schemes. The minimum required data rate of the IRs is set to $R_{\mathrm{req}}=8$ bits/s/Hz. For simplicity, we assume that $N_{\mathrm{AP}}=N_{\mathrm{PS}}$. As can be seen from Figure \ref{fig:power_vs_antennas},  the total transmit power decreases with increasing numbers of antennas. In fact, the extra degrees of freedom offered by increasing numbers of antennas can be exploited for more efficient resource allocation. Specifically, with more antennas, the direction of beamforming matrices $\mathbf{V}$ and  $\mathbf{W}_k$ can be
more accurately steered towards the AP and IR $k$, respectively, which substantially reduces the
 transmit power required in Phase I and Phase II, respectively,  for satisfying the data rate requirement.  Besides,
 the additional antennas at the AP serve as additional wireless energy collectors for energy harvesting which substantially improves the efficiency of energy harvesting at the AP.  Also,   the proposed schemes  provide  substantial power savings compared to baseline  $1$ due to the proposed optimization. Furthermore,  the performance gap between the  case of perfect hardware and the proposed schemes diminishes as the numbers of antennas equipped at the AP and PS increase, since a lower transmit power and more accurate beamforming alleviate the impact of transmit and receive HWIs, respectively.  On the other hand, a higher amount of residual energy $E_{\mathrm{res}}$ reduces the total power consumption of the system. Indeed, less PS transmit power is required for wireless charging in Phase I when the AP is equipped with a certain amount of residual energy $E_{\mathrm{res}}$. This also  substantially reduces  the power waste due to the high signal propagation loss   in wireless charging in Phase I.
\begin{figure}[t]\centering\vspace*{-8mm}
%\includegraphics[width=5 in]{time_antennas.pdf}\vspace*{-3mm}
%\caption{Average Phase I and Phase II durations versus the number of antennas equipped at the PS and AP, $N_{\mathrm{AP}},N_{\mathrm{PS}}$, for residual energy $E_{\mathrm{res}}= 1$ mW. }\label{fig:time_allocation_antennas}
\includegraphics[width=4.5 in]{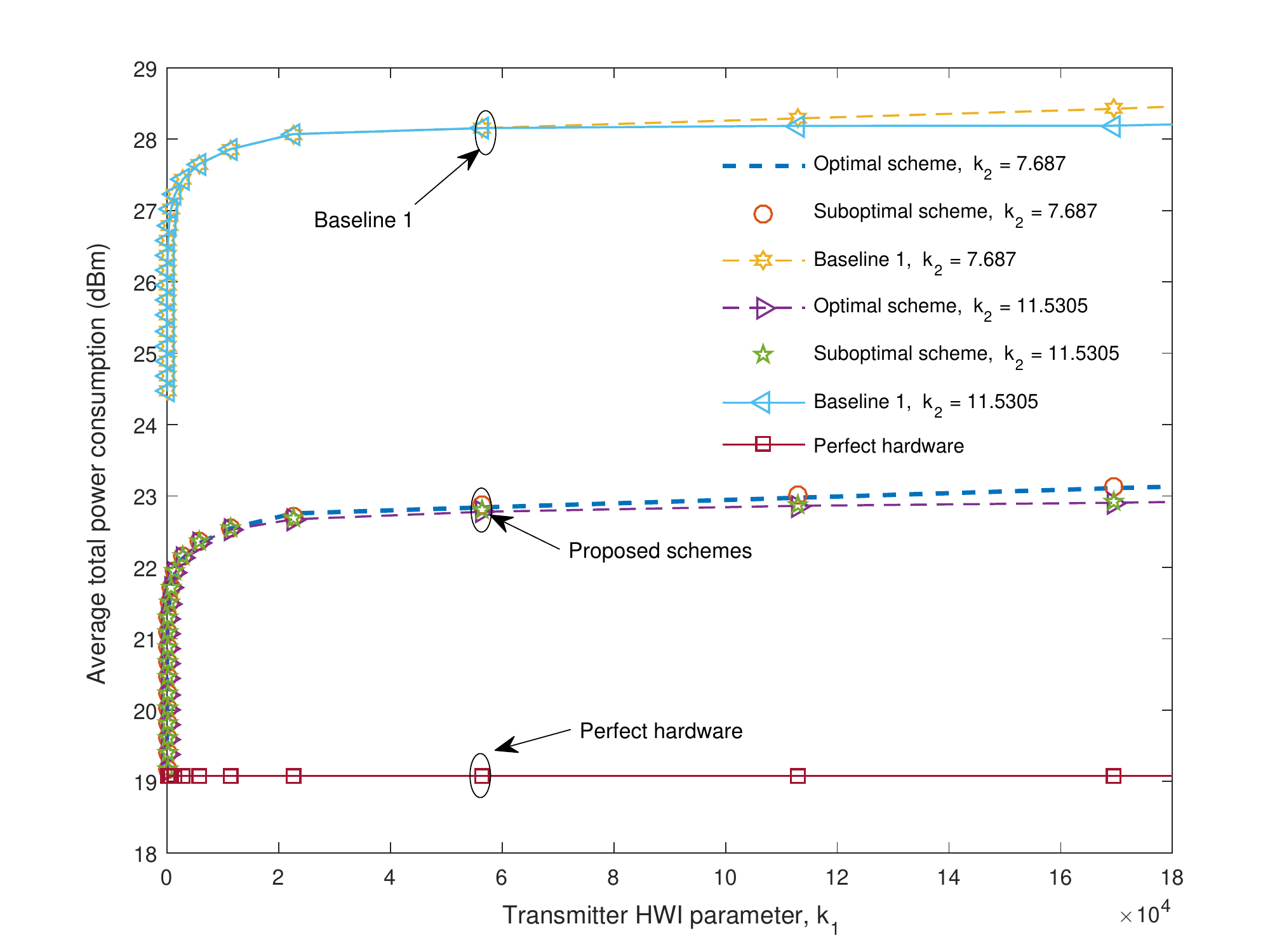}\vspace*{-4mm}
\caption{Average total power consumption (dBm)  versus the transmitter HWI parameter,   $k_{\mathrm{1}}$, for different values of $k_{\mathrm{2}}$.  } \label{fig:power_allocation_antennas}\vspace*{-4mm}\vspace*{-4mm}
\end{figure}

 Figure \ref{fig:power_allocation_antennas}
  depicts the average total power consumption
versus the transmitter HWI parameter $k_{\mathrm{1}}$ for the proposed optimal and suboptimal resource allocation schemes for different values of $k_{\mathrm{2}}$. The residual energy  and the minimum required data rate are set to $E_{\mathrm{res}}= 0$ and $R_{\min}=8$ bit/s/Hz, respectively.  As can be observed, the power consumption of the system increases with the  transmitter HWI parameter $k_{\mathrm{1}}$. This is because the power waste in \eqref{eqn:phase-PS-I}-\eqref{eqn:phase-II-AP} increases when the transmitter HWI becomes more severe leading to a less efficient resource allocation. Besides, the system power consumption of the proposed optimal and suboptimal schemes decreases with increasing  $k_{\mathrm{2}}$. As a matter of fact,
when the transmit power is properly controlled below $1$ watt, cf. \eqref{eta_func},  increasing   $k_{\mathrm{2}}$ decreases  the power waste caused by HWIs.  On the other hand, the system power consumption of the proposed schemes is significantly less than that of baseline $1$.   Besides,  for the scheme with perfect hardware,
the performance is independent of the HWI levels, of course,
and serves as an upper bound for the proposed
schemes.

\begin{figure}[t]
\centering\vspace*{-8mm}
\includegraphics[width=4.5 in]{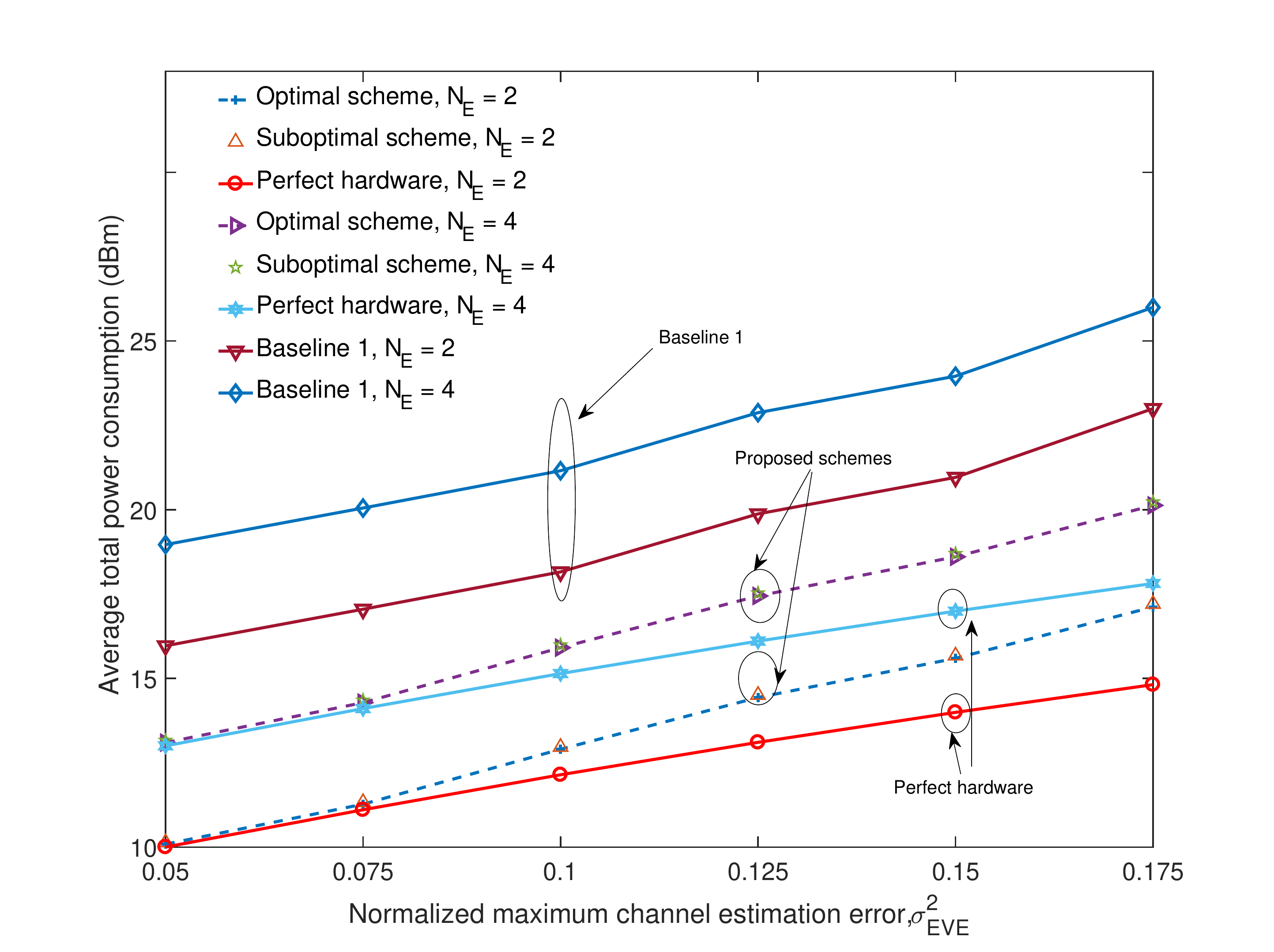}\vspace*{-4mm}
\caption{Average total power consumption (dBm) versus the normalized maximum channel estimation error for different numbers of antennas equipped at the eavesdropper, $N_{\mathrm{E}}$.}\vspace*{-8mm}
\label{fig:power_estimation_errors}
\end{figure}

In Figure \ref{fig:power_estimation_errors}, we show the average total power consumption of the system versus the normalized maximum channel estimation error.  The minimum required data rate of the IRs is set to $R_{\mathrm{req}}=4$ bits/s/Hz.  As can be observed,  the average total power consumption increases with increasing maximum channel estimation error for all  considered schemes. In fact,  as the imperfectness
of the CSI increases, both the AP and the PS become less capable of exploiting the available spatial degrees of freedom efficiently for
resource allocation. As a result, the PS has to allocate more
power to the artificial noise, $\mathbf{z}$,  to prevent interception
by the eavesdropper so as to fulfill constraint $\mathrm{C2}$. Besides, the proposed schemes  provide a substantial system power saving compared to baseline $1$, despite the imperfect CSI. Furthermore, the performance gap between the proposed schemes and the scheme with perfect hardware increases as the CSI knowledge becomes less accurate. In fact, the higher transmit power needed in the presence of imperfect CSI at both the PS and AP to fulfill the QoS requirement  worsens the impact of the HWIs. On the other hand, more power is consumed if the eavesdropper is equipped with more antennas. This is attribute to the fact that the eavesdropping capability of the eavesdropper improves with $N_{\mathrm{E}}$. To still guarantee  communication security, the PS has to allocate more power to the artificial noise  which leads to a higher system power consumption.

\section{Conclusions}\label{sec:conclude}
In this paper, we studied the power-efficient resource allocation algorithm design for providing communication secrecy in WPCNs, where we took into account a practical non-linear EH model and the residual HWIs at the transceivers. The resource allocation algorithm design was formulated as a non-convex optimization problem for the minimization of the total consumed power subject to QoS constraints at the IRs. The optimal solution of the design problem was obtained via a two-dimensional search and SDP relaxation. Besides, a low computational complexity suboptimal solution was also provided.  Numerical results demonstrated the detrimental effects of residual HWIs on performance in WPCNs. Furthermore, although the residual HWIs limit the system performance in the high transmit/receive power regimes,  the resulting performance degradation can be effectively alleviated by increasing the number of  antennas in the system and by  acquiring more accurate CSI of the eavesdropper.

\section*{Appendix - Proof of Theorem 1}
The proof is divided into two parts. In the first part, we study the rank of the information beamforming matrix, $\mathbf{W}_k,\forall k$. Then, we investigate the ranks of energy beamforming matrices $\mathbf{V}$, $\mathbf{U}$, and $\mathbf{Z}$ in the second part.  Since the SDP relaxed problem in \eqref{eqn:power_min_problem_reform2} satisfies Slater's constraint qualification and is jointly convex with respect to the optimization variables, strong duality holds and we can exploit the dual problem  \cite{book:convex} in order to study the structure of the solution of the primal problem. To this end, we write the Lagrangian function of
\eqref{eqn:power_min_problem_reform2} as:
\begin{eqnarray}\label{eqn:lagrangian}
&&\hspace*{-7mm}\mathcal{L} = (\rho_{\mathrm{AP}}\tau_{\mathrm{II}}\hspace*{-0.5mm}+\hspace*{-0.5mm}\vartheta)\Big(\sum_{k=1}^K \Tr\Big(\mathbf{W}_k\Big)\Big)\hspace*{-0.5mm}+\hspace*{-0.5mm} (\psi\hspace*{-0.5mm}+\hspace*{-0.5mm}\rho_{\mathrm{PS}}\tau_{\mathrm{I}})\Tr\Big(\mathbf{V}\Big)-\beta\Big( \Tr(\mathbf{L}^H\mathbf{V}\mathbf{L})\Big)\notag\\
&&\hspace*{-7mm}\hspace*{-0.5mm}+\hspace*{-0.5mm}\sum_{k=1}^K\lambda_k\Bigg(\sum_{j\neq k}\Tr\Big(\mathbf{H}_k\mathbf{W}_j\Big)-\frac{\Tr\Big(\mathbf{H}_k\mathbf{W}_k\Big)}{\Gamma_{\mathrm{req}}}\Bigg)-
\sum_{k=1}^K\Tr\Big(\mathbf{D}_{\mathrm{C2a}_{k}}\mathbf{S}_{\mathrm{C2a}_{k}}
          \Big({\mathbf{B}}_{\mathrm{AP}},\mathbf{U},\mathbf{W}_k,\mathbf{N},t_k\Big)\Big)\notag\\
 &&\hspace*{-7mm}-
\sum_{k=1}^K\Tr\Big(\mathbf{D}_{\mathrm{C17}_{k}}
           \mathbf{W}_k\Big)-  \Tr\Big(\mathbf{D}_{\mathrm{C}_{10}}
          \mathbf{V}\Big)   \hspace*{-0.5mm}+\hspace*{-0.5mm} \sum_{n=1}^{N_{\mathrm{AP}}} \varphi_n\Big(\sum_{k=1}^K \Tr(\mathbf{S}_n\mathbf{W}_k)\Big) \hspace*{-0.5mm}+\hspace*{-0.5mm}\sum_{k=1}^{K} \theta_k\Big( \Tr(\mathbf{H}_k\sum_{i=1}^K\mathbf{W}_i)\Big)\notag\\
         &&\hspace*{-7mm}\hspace*{-0.5mm}+\hspace*{-0.5mm}\sum_{m=1}^{N_{\mathrm{PS}}} \chi_m \Big(\Tr(\mathbf{S}_m\mathbf{V})   -b_m\Big)  +  \Delta,
\end{eqnarray}
where variables $\lambda_k$, $\beta$, $\psi$, $\vartheta$, $\varphi_n$,  $\chi_m$, and $\theta_k$  are the non-negative Lagrange multipliers associated with constraints $\widetilde{\mathrm{C1}}$, $\widetilde{\mathrm{C5}}$, $\widetilde{\mathrm{C7}}$, $\widetilde{\mathrm{C9}}$,  ${\mathrm{C12b}}$, ${\mathrm{C13b}}$, and ${\mathrm{C15b}}$, respectively. Moreover, $\mathbf{D}_{\mathrm{C2a}_{k}} \succeq \mathbf{0},\forall k$, $\mathbf{D}_{\mathrm{C}_{10}} \succeq \mathbf{0}$, and $\mathbf{D}_{\mathrm{C17}_{k}} \succeq \mathbf{0},\forall k$, are the Lagrangian multiplier matrices corresponding to constraints $\widetilde{\mathrm{C2a}}$,   $\mathrm{C10}$, and  $\mathrm{C17}$, respectively. $\Delta$ denotes the collection of terms not relevant for the proof. The Karush-Kuhn-Tucker (KKT) conditions needed for the proof are given by\footnote{We denote the optimal solution for optimization variable $x$ by $x^*$.}:
\begin{eqnarray}
\hspace*{-8mm}&&\mathbf{D}_{\mathrm{C2a}_{k}}^* \succeq \mathbf{0},\forall k,\,\mathbf{D}_{\mathrm{C10}}^* \succeq \mathbf{0}, \mathbf{D}_{\mathrm{C17}_{k}}^* \succeq \mathbf{0},\forall k,\, \\
\hspace*{-8mm}&&\beta^*, \lambda^*_k, \vartheta^*, \psi^*, \varphi_n^*, \theta_k^*,\chi_m^*\geq 0, \end{eqnarray}\begin{eqnarray}
\hspace*{-8mm}&& \mathbf{D}_{\mathrm{C17}_{k}}^* \mathbf{W}_k^* = \mathbf{0}, \forall k, \, \,\, \mathbf{D}_{\mathrm{C}_{10}}^*\mathbf{V}^* = \mathbf{0},\label{eqn:KKT_complementary}\\
\hspace*{-8mm}&& \nabla_{\mathbf{W}_k^*} \mathcal{L} = \mathbf{0}, \forall k,\,\, \,\,\,\hspace{8mm}\nabla_{\mathbf{V}^*} \mathcal{L} = \mathbf{0}. \label{KKT_cond_3}
\end{eqnarray}

Now, we investigate the rank of $\mathbf{W}_k^*$. From the complementary slackness conditions in \eqref{eqn:KKT_complementary}, we know that the columns of $\mathbf{W}_k^*$  lie in the null space of $\mathbf{D}_{\mathrm{C17}_{k}}^*$.  Hence, we focus on studying the range space of $\mathbf{D}_{\mathrm{C17}_{k}}^*$ for revealing the structure of $\mathbf{W}_k^*$. By exploiting the KKT conditions in \eqref{KKT_cond_3}, after some mathematical manipulations, we  obtain:
\begin{eqnarray}\label{eqn:matrix-equality}
\mathbf{D}_{\mathrm{C17}_k}^*
          =\underbrace{\rho_{\mathrm{AP}}\tau_{\mathrm{II}}\mathbf{I}_k +\sum_{j\neq k} \lambda_j^*\mathbf{H}_j+\frac{\mathbf{R}_{\mathbf{\widehat G}}\mathbf{D}_{\mathrm{C2a}_{k}}^*
          \mathbf{R}_{\mathbf{\widehat G}}^H}{\Gamma_{\mathrm{tol}}}+  \sum_{n=1}^{N_{\mathrm{AP}}} \varphi_n \mathbf{S}_n +\sum_{k=1}^{K} \theta_k \mathbf{H}_k}_{\mathbf{\bm{\Phi}}_k\succ\zero}-\lambda_k^*\frac{\mathbf{H}_k}{\Gamma_{\mathrm{req}_k}}.
\end{eqnarray}

Then, we study the rank of $\mathbf{D}_{\mathrm{C17}_k}^*$ for $\tau_{\mathrm{II}}>0$ by exploiting \eqref{eqn:matrix-equality} which yields:
\begin{eqnarray}
\hspace*{-5mm}&&\hspace*{5mm}\Rank\Big(\mathbf{D}_{\mathrm{C17}_k}^*
          +\lambda_k^*\frac{\mathbf{H}_k}{\Gamma_{\mathrm{req}_k}}\Big)=\Rank({\bm{\Phi}}_k)=N_{\mathrm{AP}}\\
        \hspace*{-5mm} && \Rightarrow  \Rank\Big(\mathbf{D}_{\mathrm{C17}_k}^*
          \Big)+\Rank\Big(\lambda_k^*\frac{\mathbf{H}_k}{\Gamma_{\mathrm{req}_k}}\Big)\stackrel{(a)}{\geq}\Rank\Big(\mathbf{D}_{\mathrm{C17}_k}^*
          +\lambda_k^*\frac{\mathbf{H}_k}{\Gamma_{\mathrm{req}_k}}\Big)=N_{\mathrm{AP}}\\
        \hspace*{-5mm}  &&\stackrel{(b)}{\Rightarrow} \Rank\Big(\mathbf{D}_{\mathrm{C17}_k}^*
          \Big)\geq N_{\mathrm{AP}}-1 \label{eqn:beamforming-rank},
\end{eqnarray}
where $(a)$ is due to a basic  rank inequality and $(b)$ is due to $\lambda_k^*>0$ at the optimal solution. Since $\Rank\Big(\mathbf{D}_{\mathrm{C17}_k}^*
          \Big)\geq N_{\mathrm{AP}}-1$, in order to satisfy \eqref{eqn:KKT_complementary}, it is required that  $\Rank\Big(\mathbf{W}_{k}^*
          \Big)\leq  1$. In other words, either $\mathbf{W}_{k}^*=\zero$ or  $\Rank(\mathbf{W}_k^*)=1$. On the other hand,  $\Gamma_{\mathrm{req}_k}>0,\forall k,$ and hence $\mathbf{W}_{k}^*\neq \zero$. As a result,
   $\Rank(\mathbf{W}_k^*)=1$ holds at the optimal solution of the SDP relaxed problem in \eqref{eqn:power_min_problem_reform2}. This completes the proof of the first part.

 In the second part, we show $\Rank(\mathbf{V}^*)\leq 1$. By exploiting \eqref{KKT_cond_3}, we have the following equation:
\begin{equation}\label{eqn:derivative_v}
\mathbf{D}_{\mathrm{C10}}^*
          =\rho_{\mathrm{PS}}\tau_{\mathrm{II}}\mathbf{I}_{\mathrm{PS}}+\sum_{m=1}^{N_{\mathrm{AP}}}\chi_m^* \mathbf{S}_m-\beta^*\mathbf{L}\mathbf{L}^H.
\end{equation}
Since matrix $\mathbf{D}_{\mathrm{C10}}^*
           $ is positive semidefinite, the following inequalities must  hold
\begin{eqnarray}
\lambda_{\max}\Big(\frac{\rho_{\mathrm{PS}}\tau_{\mathrm{II}}\mathbf{I}_{\mathrm{PS}}+\sum_{m=1}^{N_{\mathrm{AP}}}\chi_m^* \mathbf{S}_m}{\beta^*}\Big) \geq \lambda_{\mathbf{L}\mathbf{L}^H}^{\max} &\ge& 0,
\end{eqnarray}
where $\lambda_{\mathbf{L}\mathbf{L}^H}^{\max}$ is the  real-valued maximum eigenvalue of matrix $\mathbf{L}\mathbf{L}^H$ and $\beta^*>0$ at the optimal solution. If $\lambda_{\max}\Big(\frac{\rho_{\mathrm{PS}}\tau_{\mathrm{II}}\mathbf{I}_{\mathrm{PS}}+\sum_{m=1}^{N_{\mathrm{AP}}}\chi_m^* \mathbf{S}_m}{\beta^*} \Big)> \lambda_{\mathbf{L}\mathbf{L}^H}^{\max}$, $\mathbf{D}_{\mathrm{C10}}^*
          $ will become positive definite and a full rank matrix, i.e.,  $\Rank(\mathbf{D}_{\mathrm{C10}}^*
          )=N_{\mathrm{PS}}$. Thus, to satisfy the complementary slackness condition in \eqref{eqn:KKT_complementary}, $\mathbf{V}^*=\zero$ or   $\Rank(\mathbf{V}^*
          )\leq 1$ holds at the optimal solution\footnote{In practice, the solution of $\mathbf{V}^*=\zero $ corresponds to the case when $E_{\mathrm{res}}$ is sufficiently large and hence Phase I for wireless charging is not necessary.}.

          On the other hand, if $\lambda_{\max}\Big(\frac{\rho_{\mathrm{PS}}\tau_{\mathrm{II}}\mathbf{I}_{\mathrm{PS}}+\sum_{m=1}^{N_{\mathrm{AP}}}\chi_m^* \mathbf{S}_m}{\beta^*} \Big)= \lambda_{\mathbf{L}\mathbf{L}^H}^{\max}$,  in order to have a bounded optimal dual solution, it follows that the null space of $\mathbf{D}_{\mathrm{C10}}^*
          $ is spanned by a vector $\mathbf{u}_{\mathbf{L}\mathbf{L}^H}^{\max}$, which is the unit norm eigenvector of $\mathbf{L}\mathbf{L}^H$ associated with eigenvalue $\lambda_{\mathbf{L}\mathbf{L}^H}^{\max}$. Hence, $\Rank(\mathbf{D}_{\mathrm{C10}}^*
          )\ge N_{\mathrm{PS}}-1$. By using a similar approach as in the first part of this proof for proving $\Rank(\mathbf{W}_k^*)=1$, we can show that  $\Rank(\mathbf{V}^*)\leq 1$.

          As for proving $\Rank(\mathbf{U}^*)\leq 1$ and $\Rank(\mathbf{Z}^*)\leq 1$, we can follow the same approach as in  the second part of this proof. The details are omitted here due to page limitation. \qed


\end{document}